\documentclass[reprint,aps]{revtex4-1}
\usepackage{nccmath}
\usepackage{mathtools,amsmath,amssymb} 
\usepackage{graphicx}
\usepackage{tikz} 
\usepackage{dcolumn}
\usepackage{hyperref}
\usepackage{pifont}
\hypersetup{colorlinks=true,linkcolor=blue,filecolor=magenta,urlcolor=blue,citecolor=cyan}
\usepackage{xcolor}
\usepackage[caption=false]{subfig}

\begin{document}
\title{Soft macromolecular confinement} 
\author{Subhadip Biswas}
\email{sbiswas2@sheffield.ac.uk}
\affiliation{Department of Physics and Astronomy, University of Sheffield, Sheffield S3 7RH, UK.}
\author{Buddhapriya Chakrabarti}
\email{b.chakrabarti@sheffield.ac.uk}
\affiliation{Department of Physics and Astronomy, University of Sheffield, Sheffield S3 7RH, UK.}
\date{\today}

\begin{abstract}


We study equilibrium shapes and shape transformations of a confined semiflexible chain inside a soft lipid tubule using simulations and continuum theories. The deformed tubular shapes and chain conformations depend on the relative magnitude of their bending moduli.  We characterise the collapsed macromolecular shapes by computing statistical quantities that probe the polymer properties at small length scales and report a prolate to toroidal coil transition for stiff chains. Deformed tubular shapes, calculated using elastic theories, agree with simulations. In conjunction with scattering studies, our work may provide a mechanistic understanding of gene encapsulation in soft structures.
\end{abstract}
\maketitle

\emph{Motivation:} When a polymer is confined in a cavity, the number of configurations it can explore (in comparison to a free chain) is reduced resulting in a lower entropy state. This reduction in the chain conformational entropy manifests as a pressure exerted by the chain on the walls of the container~\cite{b:degennes1980}. Since all physical systems have finite boundaries, theoretical and experimental investigations rooted in calculating static and dynamic behaviour of confined chains has been of long standing interest~\cite{b:degennes1980, b:rubinsteincolby2003} (see  Refs [1-9] in~\cite{p:hsiaoping2013}). In recent times, theoretical~\cite{p:brochard2005, p:chen2007, p:avramova2006,p:chen2018, p:mirzaeifard2016, p:hsiaoping2013, p:odijk1983, p:sheng2001, p:milchev2011, p:cifra2009, p:kim2013, p:chen2004, p:dai2013, p:tree2013, p:ravcko2013, p:cifra2009jchemphys, p:cifra2012, p:kalb2009, b:guevorkian2009, p:huang2015, p:livadaru2003}, experimental~\cite{p:rustom2004, p:roux2001, p:karlsson2003, p:tokarz2005, p:thomas2005, p:davis2008, p:onfelt2004}, and computational~\cite{p:hsiaoping2013, p:odijk1983, p:sheng2001, p:milchev2011, p:cifra2009, p:kim2013, p:chen2004, p:dai2013, p:tree2013, p:ravcko2013, p:cifra2009jchemphys, p:cifra2012, p:kalb2009, b:guevorkian2009, p:huang2015, p:livadaru2003} studies of polymer chains confined in a cylindrical tube has garnered a lot of interest, due to a wide range of physical applications including gel permeation chromatography~\cite{p:moore1964}, oil recover \emph{etc.}~\cite{p:wever2011}. 

Most studies so far have focused on macromolecular confinement in rigid tubes \cite{p:hsiaoping2013, p:odijk1983, p:sheng2001, p:milchev2011, p:cifra2009, p:kim2013, p:chen2004, p:dai2013, p:tree2013, p:ravcko2013, p:cifra2009jchemphys, p:cifra2012, p:kalb2009, b:guevorkian2009, p:huang2015, p:livadaru2003}. In this paper we study the effect of soft confinement on the conformational properties of a semiflexible polymer. Our motivation stems from studying biological structures with modulii $\sim 10-20 k_{B} T$, which are pliant such that a confining macromolecule with dimensions larger than the confining box is able to deform it bringing about a shape change. Such deformed shapes are observed in several biophysical contexts, \emph{e.g.} during bacterial conjugation involving gene transfer between bacteria via pili~\cite{p:rustom2004,p:thomas2005}, intercellular transport of DNA, \cite{p:rustom2004} and in bioengineering applications \emph{e.g.} DNA confined in a lipid tether connecting two giant unilamellar vesicles (GUVs) \cite{p:karlsson2003, p:tokarz2005}.

\emph{Background:} The conformational properties of flexible and stiff polymer chain inside rigid channels having different cross-sections (\emph{e.g.} rectangular/cylindrical \emph{etc.}) is now well understood. Scaling and mean field theories have been extensively used to probe the static and dynamic properties of such confined polymer chains\cite{p:hsiaoping2013, p:odijk1983, p:sheng2001, p:milchev2011, p:cifra2009, p:kim2013, p:chen2004, p:dai2013, p:tree2013, p:ravcko2013, p:cifra2009jchemphys, p:cifra2012, p:kalb2009, b:guevorkian2009, p:huang2015, p:livadaru2003}, with the aim of validating the blob picture \cite{b:degennes1980}. These studies show that the extension of the polymer chain transverse to the confining direction of the tube scales with the polymer length $N$ and the undeformed tube diameter $D$, as $R_{||} \sim N D^{-\frac{2}{3}}$, in a good solvent, in excellent agreement with theoretical predictions~\cite{b:degennes1980}.

The statistical properties of macromolecules under soft confinement however, are relatively unexplored. Long semiflexible polymers confined in soft tubes the chain adopts a globular configuration if the radius of gyration of the chain in free space is greater than the diameter of the confining tube, \emph{i.e.} $R_{g} > D$~\cite{p:brochard2005}. Assuming a spherical ansatz, such a soft tubular confinement results in a chain with statistical properties intermediate between a collapsed globule, \emph{i.e.} ($R_g \sim N^{\frac{1}{3}}$), and a free polymer chain in a good solvent ($R_g \sim N^{\frac{3}{5}}$). Electrophoresis experiments of DNA confined inside a lipid tubule \cite{p:tokarz2005} reveal a ``snake'' to ``globule'' conformational transition akin to the theoretical predictions \cite{p:brochard2005}. The intensity of the fluorescent DNA, used as a marker to probe the conformational transition, shows a linear increase in intensity above a critical polymer size signalling a collapse to a globular state \cite{p:tokarz2005}. Simulations to confirm the scaling results of a polymer confined in soft tubes have been carried out using Monte Carlo \cite{p:chen2007,p:mirzaeifard2016} and molecular dynamics \cite{p:avramova2006}.
\begin{figure}
\includegraphics[width=\linewidth]{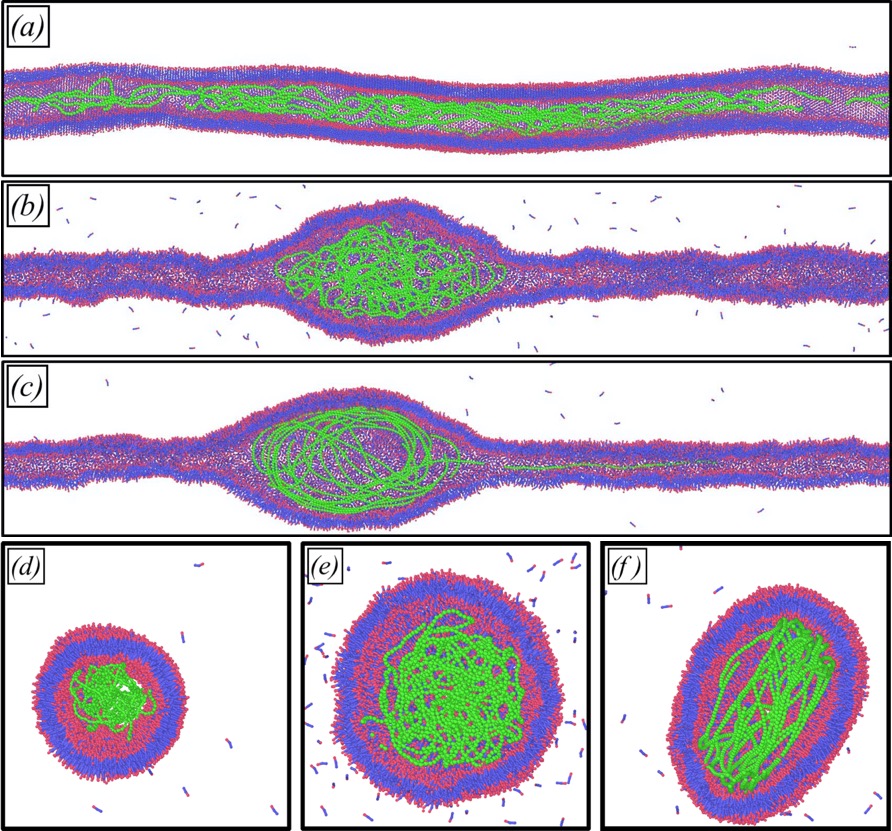}
\caption{Equilibrium snapshots of axial and radial cross section of a semiflexible polymer (green beads) of size $N=2000 \sigma$, confined in a bilayer lipid tubule with hydrophilic head groups (red beads) and hydrophobic tails (blue) obtained from CGMD simulations (SI). For rigid tubules $\kappa = 24 k_{B} T $ (a) \& (d) a swollen chain is seen, while for softer $\kappa = 12 k_B T$ tubules (b) \& (e) a globular conformation is observed. A prolate ellipsoidal conformation is observed for $l_p = 13 \sigma$, while for $l_{p} \simeq 200 \sigma$, (such that $l_{p} \gtrsim \xi$) a toroidal coil (c) \& (f) is seen. (see also SI Movies.)}%
\label{fig:swollen_globule}%
\end{figure}

In this paper we investigate the equilibrium shape of a lipid tubule with a confined semiflexible polymer using coarse-grained molecular dynamics simulations and continuum elastic theories. We characterise the confined macromolecular shapes of the chain as a function of the \emph{(i)} length of the polymer chain $N \sigma$, \emph{(ii)} its persistence length $l_{p}$, \emph{(iii)} equilibrium radius $R_{0}$, and \emph{(iv)} the bending modulus $\kappa$ of the confining tubule. For soft tubules \emph{i.e.} $\kappa \leq 15 k_{B} T$, the entropic pressure from the polymer chain is enough to deform the tubule resulting in a globular chain conformation. In this regime, the maximum radial extent $R_{\perp}$ scales with the polymer size $N \sigma$ as $R_{\perp} \sim N^{2/5} \sigma$~\cite{p:brochard2005}. However, unlike a spherically symmetric chain conformation assumed in scaling studies we observe an ellipsoidal globular shape with parts of the chain leaking into the undeformed sections of the tube \emph{i.e.} $R_{||} > R_{\perp}$ (see Fig.\ref{fig:swollen_globule}). Consequently, the bounding bilayer tubule adopts a ``drop on a fibre''~\cite{b:degennes2013} geometry with the maximum radial bulge $D_{m}/2$ showing a power law scaling with the polymer size as $\langle D_{m}/2 \rangle \sim N^{0.27 \pm 0.02}$. We compute the pressure $P$ required to bring about the volume change from a cylindrical to a prolate ellipsoidal tubule by coarse-graining the tubular profiles obtained from simulations and using a variational formulation. This entropic pressure exerted by the polymer is directly proportional to the chain size $N \sigma$, and the persistence length $l_p$ and inversely proportional to the radius of the confining tube $R_0$. We report a novel conformational transition of the confined polymer from a prolate ellipsoid to ordered toroidal coil as a function of its persistence length $l_{p}$. We identify and compute statistical measures to distinguish between different compact conformational states of the confined chain based on \emph{e.g.}, \emph{(a)} tangent-tangent correlation function, \emph{(b)} the asphericity parameters $\Delta$, and $\Xi$, \emph{(c)} radial distribution function $g(r)$, and \emph{(c)} monomer density distribution $\rho(r)$ for hard spherical, and soft tubular confinement for semiflexible chains (with different $l_p$), in good and bad solvents. It is important to note that our results are valid in the regime $\xi << L$, where $\xi$ corresponds to the deformation of the tubule in the axial direction (see Fig.~\ref{fig:swollen_globule}), and $L$ is the length of the tubule. For shorter tubules and long polymer chains, such that $\xi/L \gtrsim 1$, the integrity of the membrane is compromised resulting in a disintegration of the lipid tubule and the chain leaking out into free solution.

\emph{Tubule properties:} We consider a bilayer membrane tubule of radius $R_0$, and length $L$ having a constant area $A_0 = 2 \pi R_0 L$ with $R_0 << L$. The elastic energy of deformation of the tubule is described by the Helfrich-Canham Hamiltonian \cite{p:deuling1976}
\begin{equation} \label{1.1}
\mathcal{F} = \frac{\kappa}{2} \int_{A_0} (c_1 +c_2 -c_0)^2 dA + \Sigma \int_{A_0} dA - \Delta p \int_{V_0} dV, 
\end{equation}
where $\kappa$ and $\Sigma$ correspond to the bending modulus and the surface tension of the membrane respectively, and $\Delta p$ denotes the pressure difference acting across the membrane. The energy is expressed in terms of the two principal curvatures $c_1$, and $c_2$ while $c_0$ is the spontaneous curvature of the membrane. For a bilayer membrane as in our case $c_0 \approx 0$. Further, since we consider a closed tubule (see SI) the Gaussian curvature term in Eq.\eqref{1.1} integrates to a constant and is ignored in our analysis. The undeformed tube radius $R_0 = \sqrt {\frac{\kappa}{2 \Sigma}}$~\cite{p:derenyi2002,p:powers2002} is obtained by setting $\Delta p = 0$ in Eq.\eqref{1.1} using an area $A_0$, and volume $V_0 = \pi R^{2}_{0} L$. Variational formulations based on the Helfrich-Canham Hamiltonian and its variants~\cite{p:seifert1991, p:svetina1989}, depending on the constraints imposed by the physical situation has been used to derive shapes of closed vesicles and tubules~\cite{p:deuling1976, p:svetina1989, p:seifert1991}.

We perform coarse-grained molecular dynamics (CGMD) simulations of bilayer tubules using the LAMMPS package (SI) following the Cooke-Kremer-Deserno model~\cite{p:cooke2005,p:harmandaris2006}. Within this model we consider a tubule of length $L=300 \sigma$, and different radii $R_{0}/\sigma = 9.2, 11.2, 13.2, 15.2 $, where $\sigma$ represents the radius of a coarse-grained bead and corresponds to the unit of length in our simulations. The equilibrium radius $R_0$ is dictated by the number of assembled coarse-grained lipid molecules, and the bending modulus of the tubule $\kappa \approx 12 k_{B} T$, obtained by fitting the height-height fluctuation spectrum to a form obtained from elastic theories~\cite{p:harmandaris2006,p:fournier2007}. The surface tension $\Sigma=\kappa/2 R^{2}_{0}$, is obtained as a consistency condition from the expression of the tether radius. 

A semi-flexible polymer is placed inside the tubule and the coupled polymer-tubule system evolved in time for $t = 5 \tau_{eq}$, where $\tau_{eq} = 1 \times 10^{6} \Delta t$, is the equilibration time of the system. The equilibration time $\tau_{eq}$ is determined by monitoring the radius of gyration $R_G(t)$ of the system undergoing a swollen to globule transition as a function of time and noting the time at which the $R_G(t)$ saturates to a steady value (SI) starting from different initial configurations. The total energy of the system fluctuates around a mean value beyond $\tau_{eq}$.

Fig.~(\ref{fig:swollen_globule} a \& d) shows the axial and radial equilibrium conformation of a confined polymer in a stiff tubule with $\kappa = 24 k_B T$. In this regime, \emph{i.e.} $\kappa \gtrsim 20 k_B T$, the entropic pressure due to the polymer is not enough to deform the tube, resulting in a swollen phase. For softer tubes however, \emph{i.e.} $\kappa \lesssim 12 k_B T$ the polymer adopts a globular conformation. Fig.~(\ref{fig:swollen_globule} b \& e) shows axial and radial conformations of the polymer in this state. A prolate bulge with $R_{||} >> R_{\perp}$, is observed in this regime. The chain conformation and hence the shape of the bulge is dependent on its persistence length $l_p$. Thus for $l_p \gtrsim \xi$, a conformational transition where the chain adopts a toroidal conformation is seen in  Fig.~\ref{fig:swollen_globule} (c). The bulge shapes observed in this regime are non-prolate (Fig.~(\ref{fig:swollen_globule} (f)) with the cross-section resembling a ``hockey-puck'' like shape.   

\begin{figure}
\includegraphics[width=\linewidth]{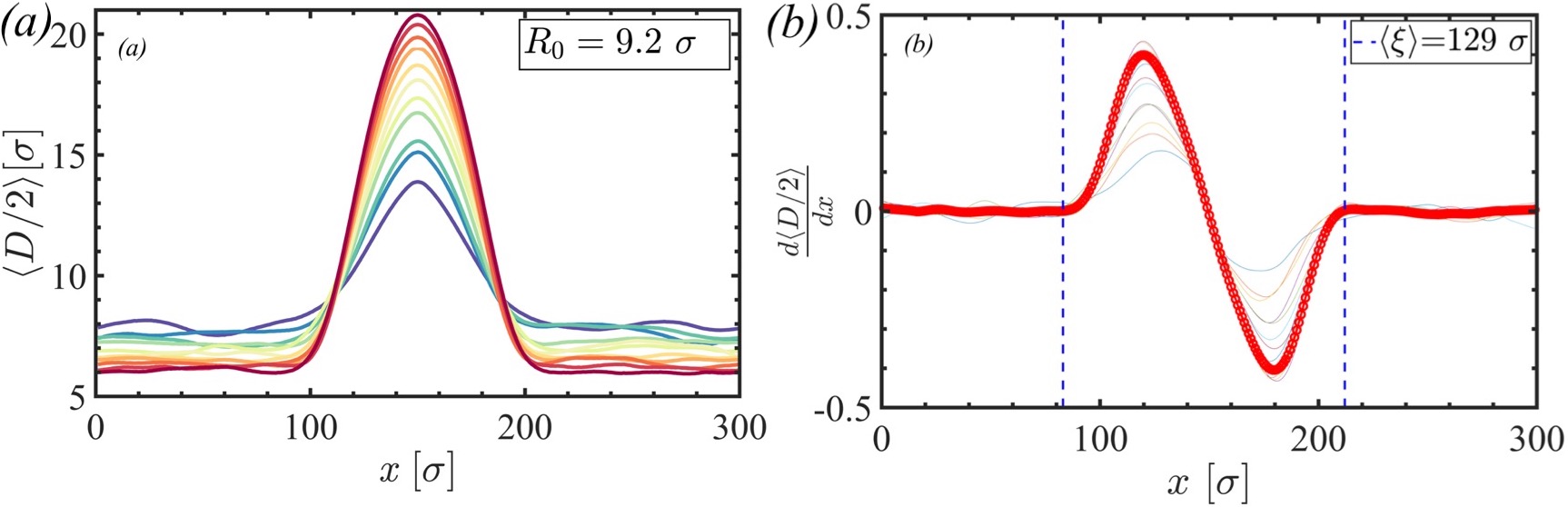}
\caption{Radial bulge $\langle D(x)/2 \rangle$ of a bilayer tubule of initial radius $R_0 = 9.2 \sigma$, along its axis $x$, encapsulating a semiflexible polymer as a function of its size $N$. The polymer size is varied between $N = N_{min} = 600$ (blue line) to $N_{max} = 2600$ (red line) in increments of $\Delta N= 200$ (a) obtained from coarse-grained simulations. The maximum radial bulge $\langle D_{m}/2 \rangle$ increases with $N$ (panel (a)), while the axial deformation length $\xi$ is independent of the chain size (panel (b)). The radius of undeformed cylindrical segments is less than $R_0$ to maintain the constant area constraint.}%
\label{fig:tubule_bulge}%
\end{figure}

The radial deformation of the axisymmetric tubule having a confined semi-flexible polymer of persistence length $l_p = 13.1 \sigma$ and different polymer lengths $N = 600, 800, \ldots 2600$ are shown in Fig.~(\ref{fig:tubule_bulge})(a). Time averaged deformation profiles \emph{i.e.} the radius of the tube $\langle D(x)/2\rangle$, as a function of the axial distance $x$ is shown, with data collected every $10^3 \Delta t$ after equilibration. While the maximum of the radial bulge $D_{m}/2$ increases with the polymer size $N$, the axial deformation length scale $\xi$ is independent of the chain length. This can be seen by taking the derivative of the bulge profile as shown in Fig.~(\ref{fig:tubule_bulge})(b). The deformed tubular shapes resemble a ``drop wetting a fibre''~\cite{b:degennes2013} (SI), with the deformation length scale in the axial direction $\xi >> D_m/2$. It should be noted that the radius of the undeformed segments is less than $R_0$ enforcing the constant area constraint during the \emph{swollen} to \emph{globule} transition. However, there is an associated increase in volume of the tubule (SI Fig. 3)). 

\begin{figure}
\includegraphics[width=\linewidth]{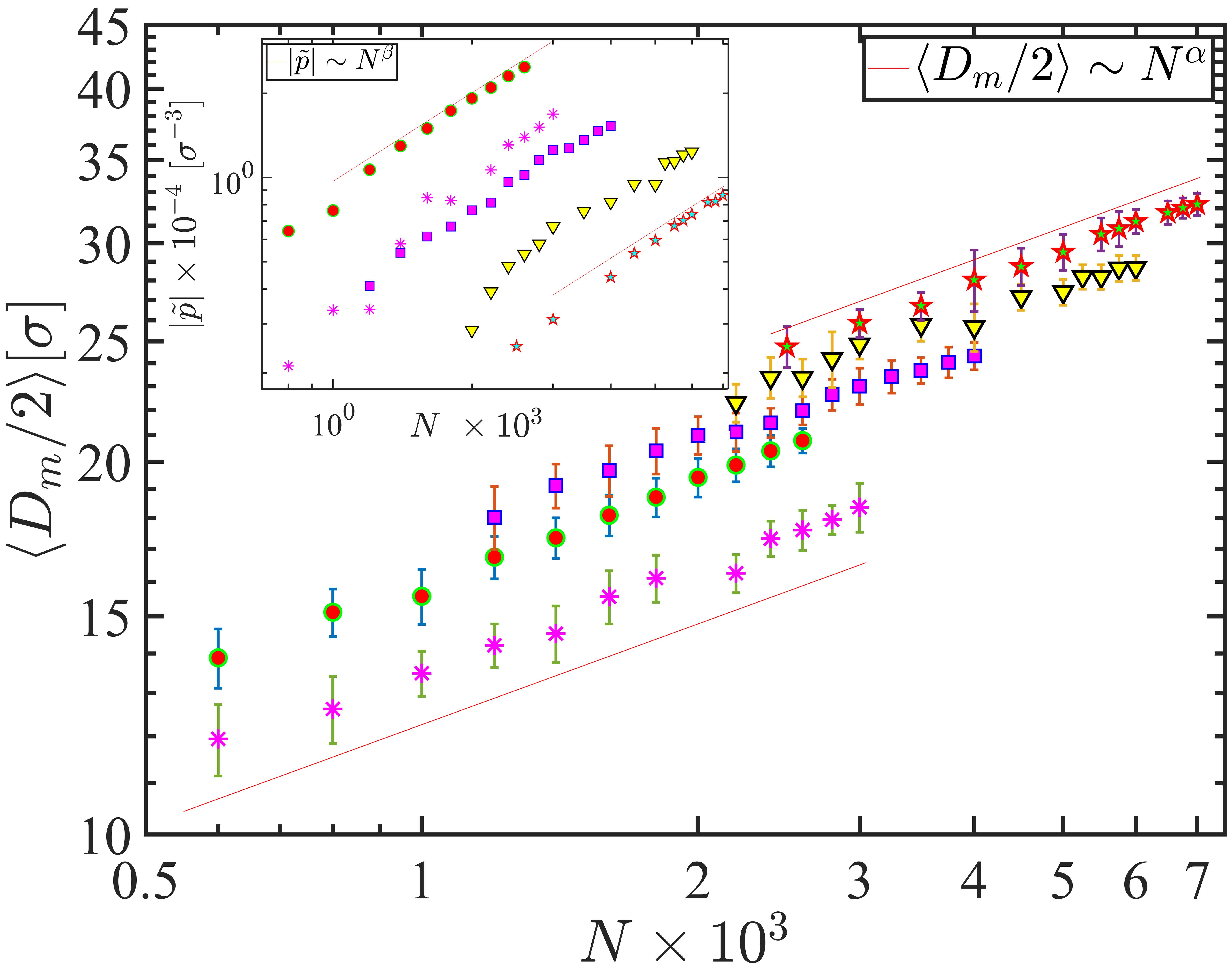}
\caption{(a.) Maximum average radial bulge $\langle D_{m}/2 \rangle$ vs.\ confined chain length $N$ for a semiflexible chain of persistence length $l_p =13.1 \sigma$ and different initial tubule radii $R_0 =  9.2 \sigma$ (\ding{108}), $11.2 \sigma$ (\ding{110}), $13.2 \sigma$ (\ding{116}), and $15.2 \sigma$ (\ding{72}) respectively. A flexible polymer in a confining tube of initial radius $R_0 = 9.2 \sigma$ (\ding{83}), is shown for comparison. Inset shows the variation of the pressure $\tilde{p}$ exerted by the polymer on the tubule as a function of its size $N$, for different initial radii $R_0$.}%
\label{fig:tube_perp_ext_pressure}%
\end{figure}

The variation of the maximum radial bulge $\langle D_{m}/2 \rangle$ of the deformed tubule as a function of the chain size $N$ ($600 \leq N \leq 7000$), for a semiflexible polymer having a persistence length $l_p=13.1\sigma$ is shown in Fig.~\eqref{fig:tube_perp_ext_pressure}. In Fig.~\eqref{fig:tube_perp_ext_pressure}, the maximum radial bulge increases with the length of the confined polymer chain, as $\langle D_m/2 \rangle \sim N^\alpha$, where $\alpha=0.27 \pm 0.02$, for tubules with initial radii $R_0 = 9.2 \sigma$ (\ding{108}) and (\ding{83}), $11.2\sigma$ (\ding{110}), $13.2 \sigma$  (\ding{116}) and $15.2 \sigma$ (\ding{72}). For a given chain length $N$, the maximum radial deformation $\langle D_m \rangle$ is less for tubes with larger initial radii $R_0$. This behaviour can be understood for flexible chains using a scaling argument. For flexible chains with excluded volume interactions, the radius of gyration scales with size as $R_G \sim N^{3/5} \sigma$. The tubule deforms when $R_G \gtrsim R_0$. This places a lower bound on the chain size $N \approx 41$ below which the tubule does not deform. The effects of semiflexibility is analysed by comparing $\langle D_{m}/2 \rangle$ against that of a flexible polymer \ding{83} in a tube of initial radius $R_0 = 9.2 \sigma$. While stiffer chains cause larger radial deformation it does not alter the scaling exponent.  

The entropic pressure arising from chain confinement $\Delta p/\kappa$ acting across the lipid bilayer is calculated as a function of the chain length $N$ using the deformed tubular shapes (SI). This is shown in the inset of Fig.~\eqref{fig:tube_perp_ext_pressure}. First, the elastic free energy difference $\Delta F$ between the reference state \emph{i.e.} a cylinder of radius $R_0$ and length $L$ ($F=0$) and the deformed state is calculated. By coarse-graining the deformed profiles, (see SI) we compute the two principal curvatures of the tubule which is then used in the expression of the Helfrich-Canham energy in Eq.~\eqref{1.1}. The free energy difference is then equated to the work done due to chain insertion in the tubule \emph{i.e.} $\Delta F = (\Delta p/\kappa) \Delta V$, where $\Delta V$ is the accompanied change in volume of the tubule. The pressure difference $\Delta p/\kappa$ scales linearly with the chain length $N$ in the strong confinement regime independent of the details of the system, \emph{e.g.} $R_0$, $l_{p}$ \emph{etc.}

\emph{Polymer properties:}
A structural characterisation of the confined chain as a function of its length $N$ allows us to estimate the chain free energy and in turn the pressure exerted by the chain on the tubule. Thus $F = - k_{B} T \ln P(R_{\perp}, R_{||})$, where $P(R_{\perp}, R_{||})$ is the probability distribution of the deformed chain in the radial and axial direction respectively. Since $R_{||}$ is independent of the chain size $N$, the probability distribution is a function of $R_{\perp}$ alone. The volume occupied by the chain $V$ can then be obtained as a function of $N$ assuming an ellipsoidal shape. Thus taking  derivative of the free energy of the chain with respect to the volume $V$, \emph{i.e.} $p = - \frac{\partial F}{\partial V}$, one can compute the pressure exerted by the chain on the tubule. For equilibrium to be established this pressure must balance the mechanical pressure difference acting across the leaflet $\Delta p/\kappa$. 

\begin{figure}
\includegraphics[width=8.8cm]{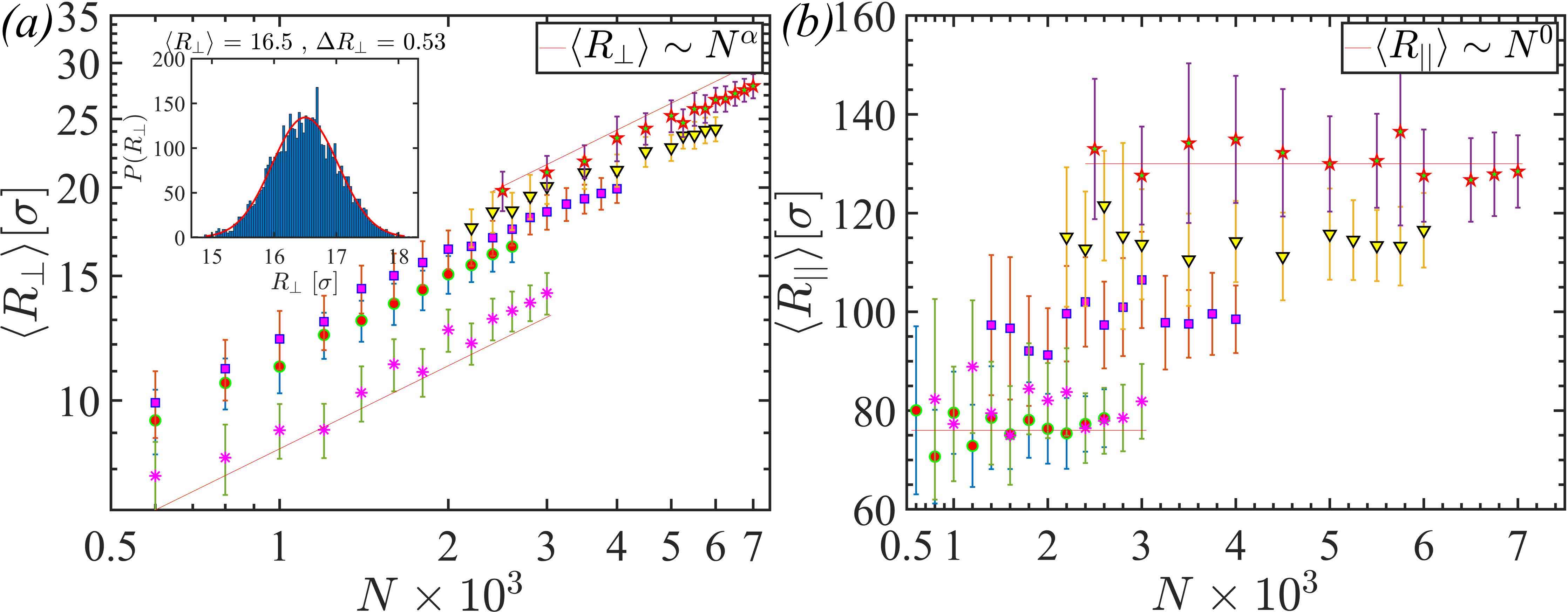}
\caption{(a) Maximum extension of the polymer along the radial direction $\langle R_{\perp} \rangle$, vs.\ chain length $N$ for initial tube radii $R_0 =  9.2 \sigma$ (\ding{108}), $11.2 \sigma$ (\ding{110}), $13.2 \sigma$ (\ding{116}), and $15.2 \sigma$ (\ding{72}) respectively. A flexible polymer in a confining tube of initial radius $R_0 = 9.2 \sigma$ (\ding{83}), is shown for comparison. Inset shows the probability distribution $P(R_{\perp})$ from which the mean extension $\langle R_{\perp} \rangle$ of the polymer is calculated. (b) Maximum extension of the polymer along the axial direction $\langle R_{||} \rangle$ is independent of confined polymer chain length $N$ for same different initial tube radii.}\label{fig:polymer_perp_radius}%
\end{figure}

Fig.~\eqref{fig:polymer_perp_radius}(a) \& (b), shows the variation of the maximum extension of a chain both perpendicular $R_{\perp}$ and parallel $R_{||}$ to the tube axis as a function of the confined chain length $N$. A distorted prolate globular conformation is observed for all values of persistence length $l_p$, and initial confining radius $R_0$. The radial extension scales with the chain length $\langle R_{\perp} \rangle \sim N^{0.38\pm0.03}$ in agreement with mean field theories~\cite{p:brochard2005}. The maximum chain extension along the axial direction $R_{||}$ is independent of the chain length $N$ in this regime. In contrast, for chains confined inside rigid tubes $R_{||}$ increases linearly with chain size. For $R_0 \gtrsim 15 \sigma$ and polymers with $N \lesssim 2000$, the chain adopts an extended conformation along the tube axis. 

While the large scale properties of confined polymer chains can be investigated using statistical signatures \emph{e.g.} $R_{\perp}$ and $R_{||}$, it does not provide information about shape transformations on length scales smaller than the bulge deformation scale $\xi$. In this regime, we probe the short distance properties of the chain conformations by computing \emph{(a)} tangent-tangent correlation function $\langle \vec{t}_{i} \cdot \vec{t}_{j} \rangle$, \emph{(b)} the asphericity parameters $\Delta$, and $\Xi$, \emph{(c)} the radial distribution function $g(r)$ (SI), and \emph{(d)} the mass distribution $\rho_{r}$ of the confined polymer. 

Persistence length of semiflexible chains under confinement has been measured in experiments~\cite{p:koster2005} via tangent-vector correlations $\langle \vec{t_i} \cdot \vec{t_j} \rangle$ along the chain backbone. Geometric confinement renormalizes the persistence length of the chain. For stiffer chains, the effective persistence length $l_e$, is smaller than the bare value $l_p = 2 \kappa_p /k_{B} T$, where $\kappa_p$ is the bending modulus of the chain~\cite{b:rubinsteincolby2003}. The effective persistence length $l_e$ depends on the ratio of the bending modulii of the membrane $\kappa$, and the polymer $\kappa_{p}$ for soft confinement, and involves the confinement length scale for hard confinement.

The tangent-tangent correlations of a semiflexible under hard spherical confinement is approximately~\cite{p:liu2008} 
\begin{equation}\label{eq:2}
\langle \vec{t_i} \cdot \vec{t_j} \rangle \approx e^{-\frac{s}{l_e}} \left[\cos\left(\frac{s}{R}\right) - 2 \frac{R}{l_e}\sin\left(\frac{s}{R}\right)-\left(\frac{R}{l_e}\right)^2\cos\left(\frac{s}{R}\right)\right].
\end{equation}

 The characteristic oscillatory behaviour of the correlation function observed in experiments is captured in this model. The leading order contribution to the correlation function in the limit $\frac{l_e}{R_0} >> 1$ is of the form  $e^{-\frac{s}{l_e}}\cos\left(\frac{s}{R}\right)$, where $R_0$ is the radius of the hard spherical confinement, and can be used to compare against simulation results of chains under soft tubular confinement. 

Stiff chains having persistence length larger than the tubule length $l_p >> L$, undergoes a conformational transition where the chain wraps the inner leaflet of confining bilayer tubule to minimize the bending energy cost. Thus, the polymer density is high near the wall and nearly zero density at the central region of the confining tubule. We characterise this shape transition by calculating the asphericity parameter $\Delta = \dfrac{3}{2}\frac{\text{Tr}~\hat{Q}^2}{(\text{Tr}~Q)^2} = 1 - 3\frac{\lambda_1 \lambda_2 +\lambda_2 \lambda_3 +\lambda_3 \lambda_1}{(\lambda_1 +\lambda_2 +\lambda_3)^2}$, expressed in terms of the variance of the eigenvalues of the traceless gyration tensor about the centre of mass, $\hat{Q}_{\alpha \beta} = Q_{\alpha \beta} - \frac{1}{3}\text{Tr}(S)E$, with $E$ unit tensor \cite{p:arkin2013}. The gyration tensor, $Q_{\alpha \beta} \equiv \langle(r_{\alpha} - \langle r_{\alpha} \rangle) (r_{\beta} - \langle r_{\beta} \rangle)\rangle = \frac{1}{N} \sum_i (r_{i}^{\alpha} - r_{cm}^{\alpha})(r_{i}^{\beta} - r_{cm}^{\beta})$, where $\alpha$ and $\beta$ indicate components of the position vector of the $i$-th monomer~\cite{p:alim2007,p:aronovitz1986,p:cannon1991,p:ostermeir2010,p:huang2019}. 

The asphericity parameter $\Delta$ lies in the range $0 \leq \Delta \leq 1$, with $\Delta = 0$ for spheres, and $\Delta = 1$ for rigid rods. The difference between oblate and prolate shape is given by the parameter
\begin{equation}\label{eq:12}
\Xi =\frac{4\text{Det}~\hat{Q}}{(\frac{2}{3}\text{Tr}~Q^2)^{\frac{3}{2}}} = \frac{4(\lambda_1 - \bar{\lambda})(\lambda_2 - \bar{\lambda})(\lambda_3 - \bar{\lambda})}{(\frac{2}{3} \sum_{i = 1}^{3}(\lambda_i - \bar{\lambda})^2)^{\frac{3}{2}}},
\end{equation}
where $\bar{\lambda} = \frac{1}{3} \sum_{i=1}^{3}\lambda_{i}$ represents the average chain size. 
\begin{figure*}
\includegraphics[width=\textwidth]{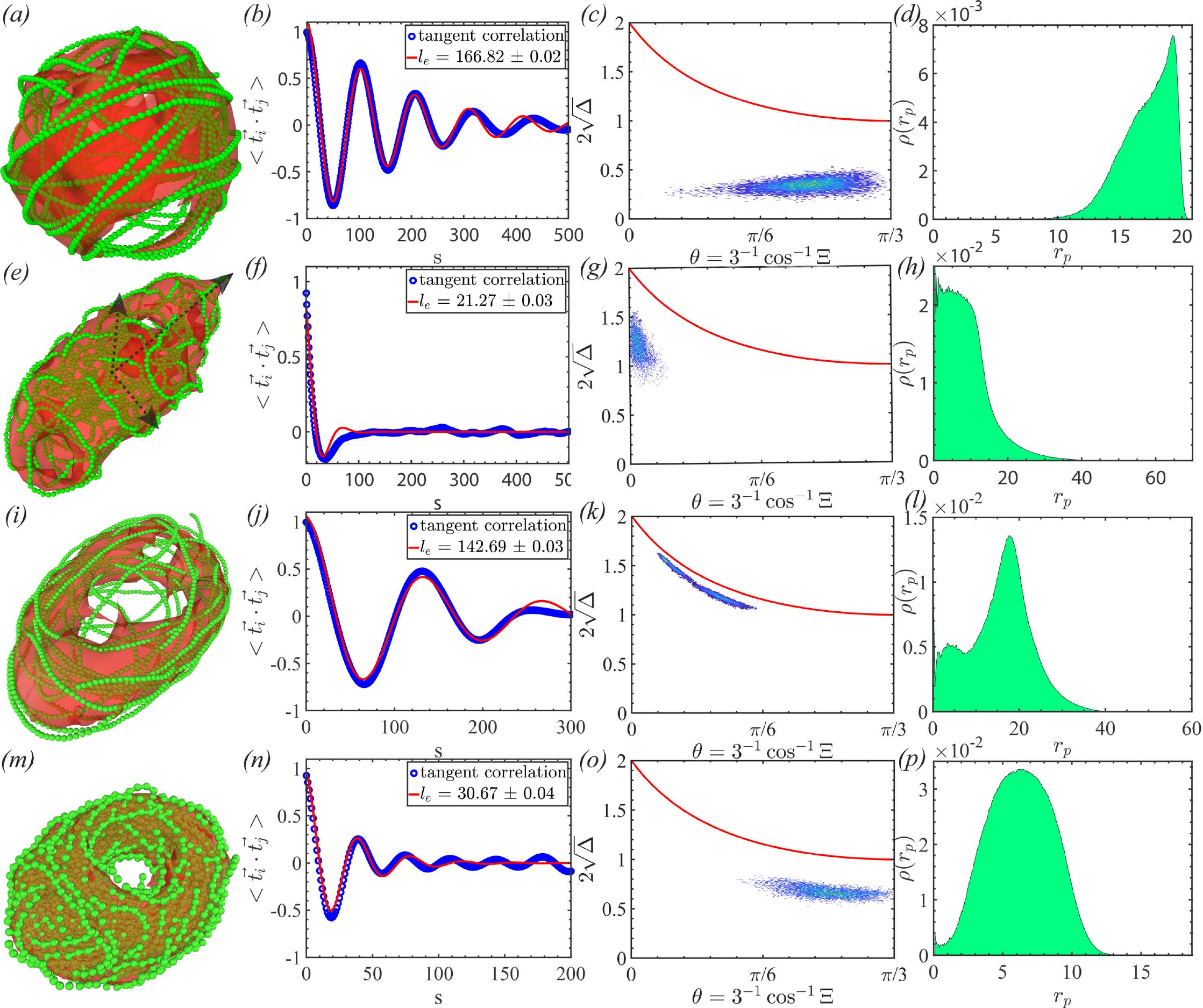}
\caption{Figure shows conformations of semiflexible polymers (a) inside a rigid hollow sphere, (e) in a soft tube $l_p = 20 \sigma$; (i) a stiffer polymer chain in a soft tube $l_p = 200 \sigma$, and (m) semiflexible polymer in a soft tube in a bad solvent. Second column (b), (f), (j), and (n) shows tangent-tangent correlation of the polymer chain. Confinement changes the effective persistence length of the chain. For stiff chains $l_p = 200 \sigma$, a conformational transition from ellipsoidal to toroidal shapes is observed, characterised by the asphericity parameters, $\Delta$, and $\Sigma$. Distribution of monomers from the centre of mass of the chain (panels (d), (h), (l), and (p)), probing short distance properties of confined chains shows differences between different confining geometries and polymer and solvent qualities.}
\label{fig:table_figures}%
\end{figure*}
The shape characterisation of confined compact structures is shown in Fig.~\eqref{fig:swollen_globule}. A space filling picture for four different compact structures corresponding to (a) spherical, (e) prolate, (i) toroidal coil and (m) collapsed prolate shapes respectively. The difference between these compact shapes are characterised in terms of the short distance properties of the chain. Panels (b), (f), (j) and (n) shows the tangent-tangent correlation function $\langle \vec \vec{t}_{i} \cdot \vec{t}_{j} \rangle$ (blue) and a two parameter fit to the leading order behavior obtained for spherical confinement (red) of the form $e^{-x/l_e} \cos(x/R)$. The effective persistence length $l_e$ of the chain under confinement, and the radius of the confining boundary $R$ is obtained from this fit. The radius of the confining sphere is $R_0 = 20 \sigma$, and for the tubule $R_0 = 9.2 \sigma$. We demonstrate the renormalisation of the persistence lengths of chains under confinement. For example the tangent-tangent correlation of a semiflexible polymer $l_p \sim 200 \sigma$, confined in a hollow sphere with rigid walls adopts a conformation such that the tangent vectors are decorrelated over $\pi R \approx 62 \sigma$, \emph{i.e.} a semicircular arc traced by the polymer inside the sphere. Consequently, the effective persistence length is renormalised to a value $l_e \sim 167 \sigma$, obtained by fitting the tangent-tangent correlation function computed from simulations to  Eq.~\eqref{eq:2}. The length scale over which tangent vectors would decorrelate if the polymer was confined inside the sphere is expected to be $\sim 2 R \approx 40 \sigma$. However, as seen in panel (i) a semiflexible polymer with $l_p = 200 \sigma$ forms a toroidal coil inside the tube. One can estimate the effective persistence length by noting that antipodal points on the ellipsoid on the equatorial plane formed by the major and minor axis is given by $\pi \sqrt{\frac{a^{2} + b^{2}}{2}}$. The magnitude of the eigenvalues of the radius of gyration tensor $R_{g}$ gives us an estimate of these length scales. Thus $a/\sigma \sim 225$ and $b/\sigma \sim 7.5$ yields an effective persistence length $l_e \sim 160\sigma$ roughly agreeing with those observed in simulations. Panel (e) represents a less stiff chain with bare persistence length $l_p \sim 20 \sigma$. In this case the effective lengthscale increases for soft confinement. In the presence of attractive interactions among the beads (\emph{i.e.} mimicking a bad solvent) a collapsed configuration is observed. While for some realisations a toroidal chain morphology is seen, as in panel (m) the short distance properties of such a polymer is drastically different from that of the toroidal chain morphology seen in panel (i). This is also borne out in the measurement of the radial distribution function $g(r)$ (SI). 

The shape transition can be quantified further using the asphericity $\Delta$ and nature of asphericity $\Xi$. As shown in panels (c), (g), (k), and (o) respectively, the heat map obtained for different configurations in the $\Delta$-$\Xi$ phase plane~\cite{p:alim2007} is different for different shapes. The red line in the figures demarcate the region between accessible and inaccessible closed shapes, while the heat map provides a measure of prolate, and oblate polymer conformations. The phase plane analysis is able to differentiate between ellipsoidal shapes in panel (g) and (k) of Fig.~\eqref{fig:table_figures}. The radial density distribution function $g(r)$ (SI) can also be used to distinguish between polymer shapes under soft confinement. The distribution of monomers inside the tube can be used to distinguish between the different conformations. Thus, from the local density of monomers, defined by $\rho(\vec{r}) = \langle \sum^{N}_{i=0} \vec{r} - \vec{r}_{i} \rangle$, we compute the monomer distribution given by $\rho(\vert \vec{r_p} \vert) = 1/N \int^{r_p}_{0} \rho(\vert \vec{r} \vert) d\vec{r}$, as a function of $r_p$. While a monomer distribution with a peak close to $r_p \sim 0$ is observed for ellipsoidal geometries, the monomer distribution is peaked about $r_p \sim 20 \sigma$ corresponding to the maximum value of the radial bulge. A similar shift in the peak of the distribution arises for collapsed chains under soft confinement, though the shape of these distributions are different. These statistical measures provides a quantitative benchmarking scheme by which conformational properties of semiflexible polymers under soft-confinement may be characterised. 

{\emph{Discussion}:} We have developed a multi-scale model to compute statistical properties of polymer chains under soft confinement and characterise the confined macromolecular phases as a function of the polymer chain length $N$, stiffness $l_p$, radius of the confining tube $R_0$ and its bending modulus $\kappa$ and report a swollen to globular conformational transition as the chain size $N$ and stiffness $l_p$ is increased. The maximum radial bulge of the polymer $\langle R_{\perp} \rangle$ scales with the chain size as $R_{\perp} \sim N^{0.38 \pm 0.03}$ in agreement with scaling theories~\cite{p:brochard2005}. The statistical properties of the tubule however, follow a different scaling behaviour than that of the polymer with the maximal radial bulge $\langle D_{m}/2 \rangle$ increasing with the polymer size $N$ as $\langle D_m/2 \rangle \sim N^{0.27 \pm 0.02}$. We have developed statistical measures to characterise shapes of confined polymers in soft tubules and their associated conformational transitions by computing statistical properties of chains at short distances, which cannot be captured in mean-field/scaling studies. In this context, we report a novel ellipsoid to toroidal chain conformational transition as a function of the persistence length $l_p$. Further, we have developed a continuum description to predict the equilibrium shapes of deformed tubules and in turn the macroscopic size of the confined semiflexible polymer using a variational formulation (SI). By parameterising the deformed tubular shapes using an ansatz, we compute the principal curvatures of the tubular membrane and hence the free energy of the tubule using the Helfrich Hamiltonian\cite{p:deuling1976}. By computing the volume change between the undeformed cylindrical tubule and the ellipsoidal globule we calculate the pressure exerted by the polymer chain on the walls of the tube (SI). This pressure can also be obtained by computing the derivative of the chain free energy with respect to the volume occupied by the confined chain, numerically, \emph{i.e.} $p = - \frac{\Delta F}{\Delta V} \vert_{N, T}$. Due to the limitations in the chain sizes, $N$, we have not explored this in our present study. 

Biological function crucially relies on a tight regulation between confinement and deconfinement of biomacromolecules in soft structures and their transport through them for suitable function, e.g. chromatin compaction inside the cell nucleus~\cite{p:maji2020}, transport of mRNA through the NPC nuclear pore complex~\cite{p:abhishek2020}, transfer of bacterial genome through pili during conjucation~\cite{p:rustom2004,p:thomas2005} \emph{etc.}. Since the probability distribution of polymer shapes provides a measure of the entropy of the confined chains, a change in this distribution provides a way to compute the equilibrium thermodynamic behaviour of such confined polymers. Shape transformations between different conformations can therefore be linked to associated changes in free energies. Thus dynamics of polymer conformational transitions and transport under soft confinement can be probed by, \emph{(i)} following the chain conformations using coarse-grained molecular dynamics simulations at the mesoscale and \emph{(ii)} using statistical measures of conformational shapes (see Fig.~\eqref{fig:polymer_perp_radius}(a)) to compute the free energies of chains and develop transition state type models at the macroscale~\cite{p:biswas2020}. Scattering studies based on fluorescent labelling of monomers might be able to probe conformational properties of semiflexible chains under soft confinement. We believe that our theoretical work will motivate such experimental studies in this direction which will in turn shed light on biophysical problems involving soft macromolecular confinement.


{\emph{Author Contributions}}: BC designed the research and obtained funding. SB carried out the research and analysed the data. BC supervised the research. SB and BC wrote the paper.

{\emph{Acknowledgements}:}
SB and BC thank University of Sheffield, IMAGINE: Imaging Life grant for financial support. We thank Dr Biswaroop Mukherjee, and others in the Chakrabarti lab for a critical reading of our manuscript.


\newpage
\onecolumngrid
\begin{center}
\LARGE{Supplementary Information}
\end{center}
\setcounter{figure}{0}  
\setcounter{equation}{0}  
\section{Coarse Grained Simulations:}

\emph{Description of the Model:}

To probe the conformational properties of a homopolymer chain confined in a soft tube, we perform Langevin dynamics simulation to verify theoretical results obtained from continuum theories. For mesoscopic simulation of the membrane tubule, we adopt coarse-grained \emph{Cooke-Kremer-Deserno} lipid bilayer model \cite{p:cooke2005}. A coarse grained three bead model of lipid molecules is represented by one hydrophilic head bead followed by two hydrophobic tail beads. In addition to a bonded interaction, head and tail beads interact via a repulsive Weeks-Chandler-Anderson (WCA) or truncated Lennard-Jones(LJ) potential
\begin{eqnarray}\label{eq:1.1}
V_{\text{rep}}(r) =  
\begin{cases}
4 \epsilon  \left[\left(\frac{b}{r}\right)^{12} -\left(\frac{b}{r}\right)^{6} +\frac{1}{4}\right],~~~~~r_c \geq r ,\\
0, ~~~~~~~~~~~~~~~~~~~~~~~~~~~~~~~~~ r > r_c,
\end{cases}
\end{eqnarray}
where $\epsilon$ is the potential well depth and $r_c = 2^{\frac{1}{6}}b$ is the location of the minima. The geometric parameters of the three bead lipid model with $b_{\text{head,head}} = b_{\text{head,tail}} = 0.95 \sigma$ and $b_{\text{tail,tail}} = \sigma$ is such that the tail beads are slightly bigger than the head bead. To mimic fluidic behaviour of the membrane non-bonded interactions between coarse-grained (CG) tail beads is incorporated. Thus, the effective hydrophobic interactions between tail beads mimicking the implicit solvent is given by
\begin{eqnarray}\label{eq:1.2}
	V_{\cos}(r) =  
	\begin{cases}
		-\epsilon,~~~~~~~~~~~~~~~~~~~~~~~~~r<r_c,\\
		-\epsilon \cos ^2 \left(\frac{\pi(r-r_c)}{2 w_c}\right),~~~~~r_c +w_c \geq r \geq r_{c},\\
		0, ~~~~~~~~~~~~~~~~~~~~~~~~~~~r > r_{c} + w_{c},
	\end{cases}
\end{eqnarray}
an attractive potential of depth $\epsilon$ with the potential decaying to zero smoothly in the interval between $r_C$ and $r_{c} + w_{c}$. The width of the attractive region is $w_{c}$. The bonded interaction between tail beads and head bead is modeled by the finite extensible nonlinear elastic (FENE) potential,
\begin{eqnarray}\label{eq:1.3}
	V_{\text{FENE}}(r) =  - \frac{k_{b} r_{0}^2}{2} \log\left[ 1- \left(\frac{r}{r_{0}}\right)^2\right].
\end{eqnarray}
Here $r_0$ is the maximum extension of the spring.

A bond bending potential, that generates a straight conformation of lipid molecules is given by
\begin{eqnarray}\label{eq:1.4}
	V_{\text{angle}} =  \frac{k_{\theta}}{2} \left[ \theta - \theta_0 \right]^2,
\end{eqnarray}
where equilibrium angle $\theta_0$ between three beads is $\theta_0 = 180^{\circ}$ and $k_{\theta}$ is the spring constant corresponding to the bond-bending potential. For lipid interactions $k_{\theta} = 10 \epsilon/\sigma^{2}$, while for the polymer $k_{\theta}$ ranges between $10 \epsilon/\sigma^{2}$ and $100 \epsilon/\sigma^{2}$ depending on the semiflexibility of the chain.

A \emph{Kremer-Grest} bead-spring model has been used to simulate CG polymer of $N$ monomers connected by FENE springs as shown in Eq.~\eqref{eq:1.1}. Persistence length of the polymer depends upon the equilibrium bond angle $\theta_0$ between three consecutive beads and the spring constant of the angle potential $k_{\theta}$ as described in Eq.~\eqref{eq:1.4}. Monomers of diameter $\sigma$ with bond cutoff length $r_{0} = 1.5 \sigma$ which feels repulsive interaction with other monomers and lipid molecules due to WCA potential in Eq.~\eqref{eq:1.1}.

We perform molecular dynamics simulations of polymer-membrane system using LAMMPS~\cite{lammps} with a Langevin thermostat. A random polymer chain is inserted inside a cylindrical bilayer membrane (initially both leaflet having the same number of lipids) as a starting configuration. We minimise the energy of the compound polymer-membrane system via a conjugate gradient minimisation scheme. To obtain equilibrium conformations for both polymer and membrane we perform simulations in a constant volume, for $t = 5 \times 10^{6} ~\Delta t$, where $\Delta t = 10^{-2} \tau_{LJ}$  (where $\tau_{LJ} =  \sqrt{\frac{m \sigma^2}{\epsilon}}$) being the time step of the simulation. To reach the equilibrium state from its initial configuration (where the polymer is in swollen state) it takes $t_{eq} \sim 2.5 \times 10^{6} \Delta t$. 

The radius of gyration $R_G$ and the radial $R_{\perp}$ and axial extension $R_{||}$ of the polymer is monitored as a function of time. For chains undergoing a conformational collapse the $R_{G}$ decreases as a function of time. Equilibration time $\tau_{eq}$ is obtained by noting the time scale at which $R_{G}$ saturates to a steady value. The total energy of the system of the system in this phase fluctuates around a fixed value. For polymers that remain in the swollen phase, we collect data for $t >> \tau_{eq}$ obtained from the collapse runs. 

Random initial polymer conformations are generated using a MATLAB script. With this model canonical self-assembly of the flat membrane is observed in simulations, however, spontaneous self-assembly of bilayers to form a tubule is not possible~\cite{p:cooke2005}. Therefore, we generate a cylindrical bilayer tubule as our initial configuration and minimize the total energy using LAMMPS to determine equilibrium tubular shape fluctuations. 

\section{Data Analysis:}

\subsection{Quantifying conformational transitions}
\emph{Coil to globule} transition:

\begin{figure}[!h]
\centering
\includegraphics[width =0.7\textwidth]{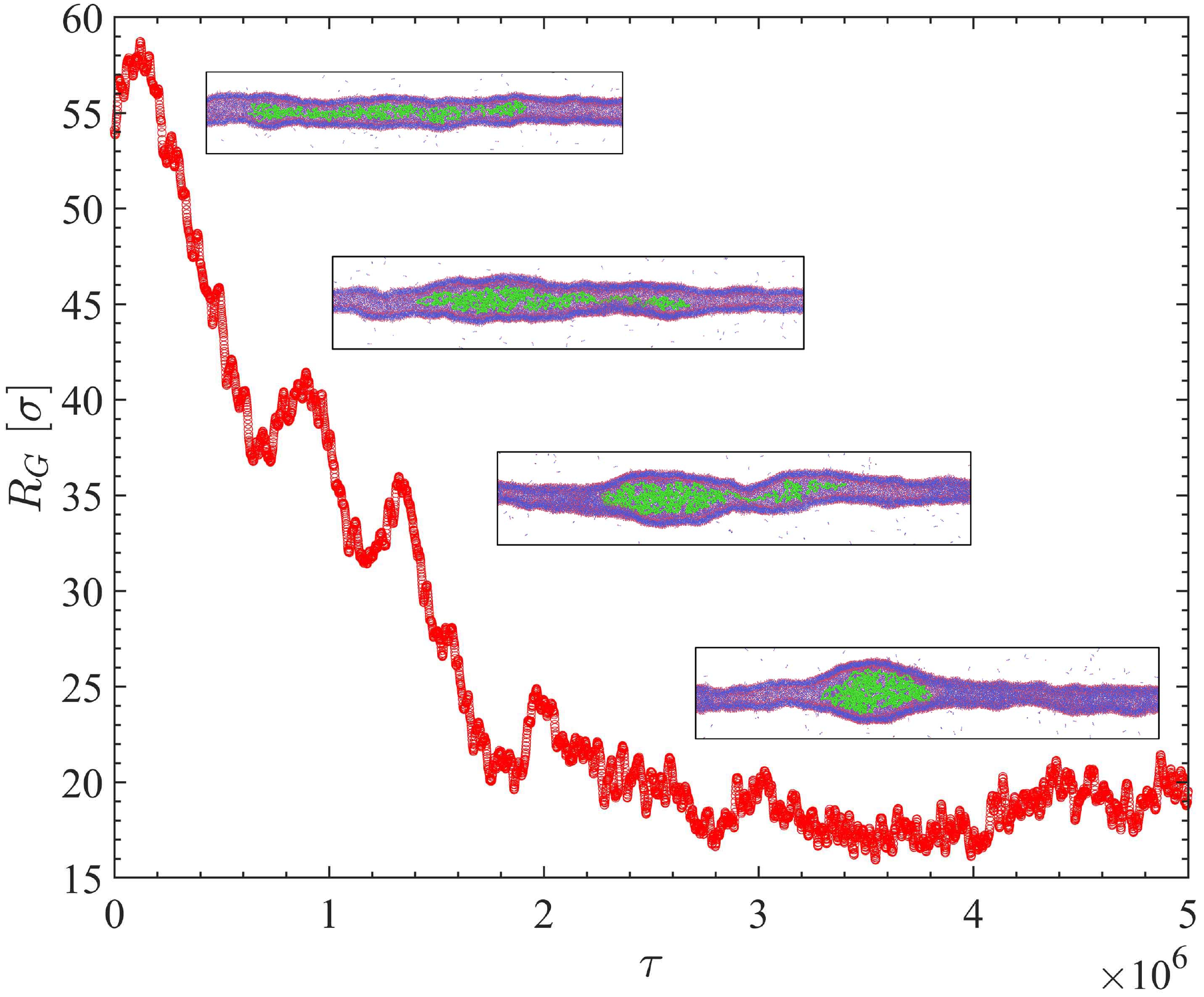}
\caption{\emph{coil to globule} transition as function of time. Radius of gyration $R_G$ of the polymer chain is changing from its higher value to an equilibrium value.}
\label{fig1}
\end{figure}

A semiflexible polymer chain inside a tube whose bending modulus $\kappa >> k_{B} T$, is rigid, in which the polymer adopts an elongated conformation. On the contrary, if $\kappa \sim k_{B} T$, polymer can deform the tube and adopt a globular structure. Depending upon the bending rigidity of the polymer chain, its size, bending modulus of the tubule, membrane tension \emph{etc.}, the confined polymer and the tubule can have different shapes. For a semiflexible chain with, $l_p >> R_{0}$ ($R_{0}$ is the radius of the confining tube), the chain adopts a toroidal shape, in order to minimize its bending energy. The resulting shapes are qualitatively different than those obtained for softer chains, \emph{i.e.} $l_p \sim D$. In the case, $l_p \sim D$, polymer elongated along the tube axis and eigenvalues of other two orthogonal principal axes are  similar. Prolate shapes obtained in this regime is shown in the main text Fig.~(6). 

As shown in the Fig.~\eqref{fig1}, a polymer chain placed in an extended state is placed in a cylindrical bilayer membrane tubule. Initial radius of gyration of the polymer chain of size $N = 2000$ is $R_G \sim 58 \sigma$. The time trace shows $R_{G}(t)$ decreasing as a function of time indicating a collapse transition to a globular state. Configuration snapshots indicate a transition to a final ellipsoidal globular structure. We generate the data in two steps; the polymer-tubule system is simulated \emph{(i)} till the formation of a globule \emph{i.e.} $\tau = 5 \times 10^{6} \Delta t$, and thereafter \emph{(ii)} the final globular configuration from \emph{(i)} are further simulated for a further $\tau = 5 \times 10^6$ time steps in a constant volume ensemble with a constant temperature enforced via a Langevin thermostat. Our polymer conformational statistics data is collected in the equilibrium state in intervals of $\tau = 10^{3} \Delta t$.

\begin{figure}[!h]
\centering
\includegraphics[width=.33\textwidth]{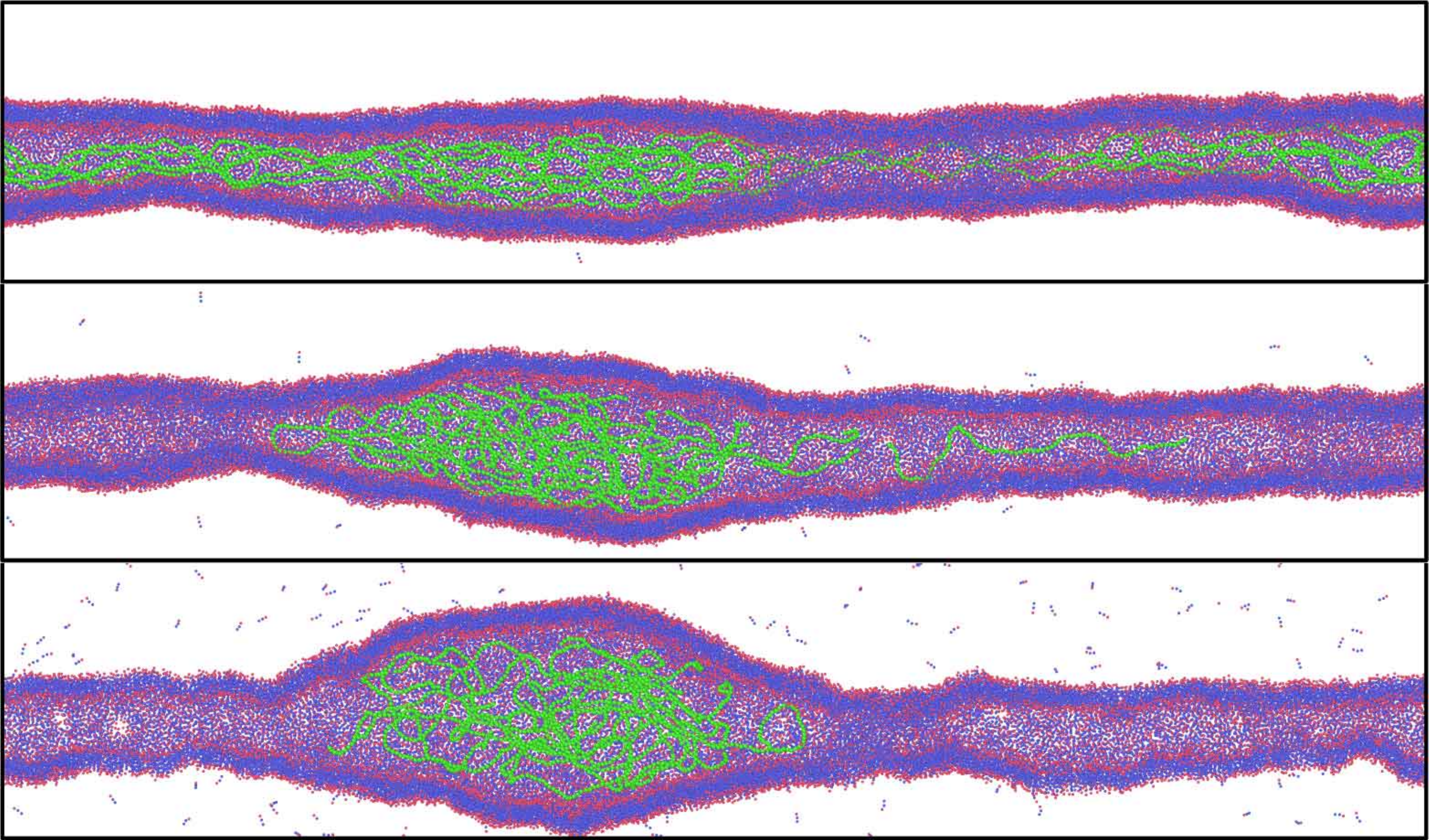}\hfill%
\includegraphics[width=.33\textwidth]{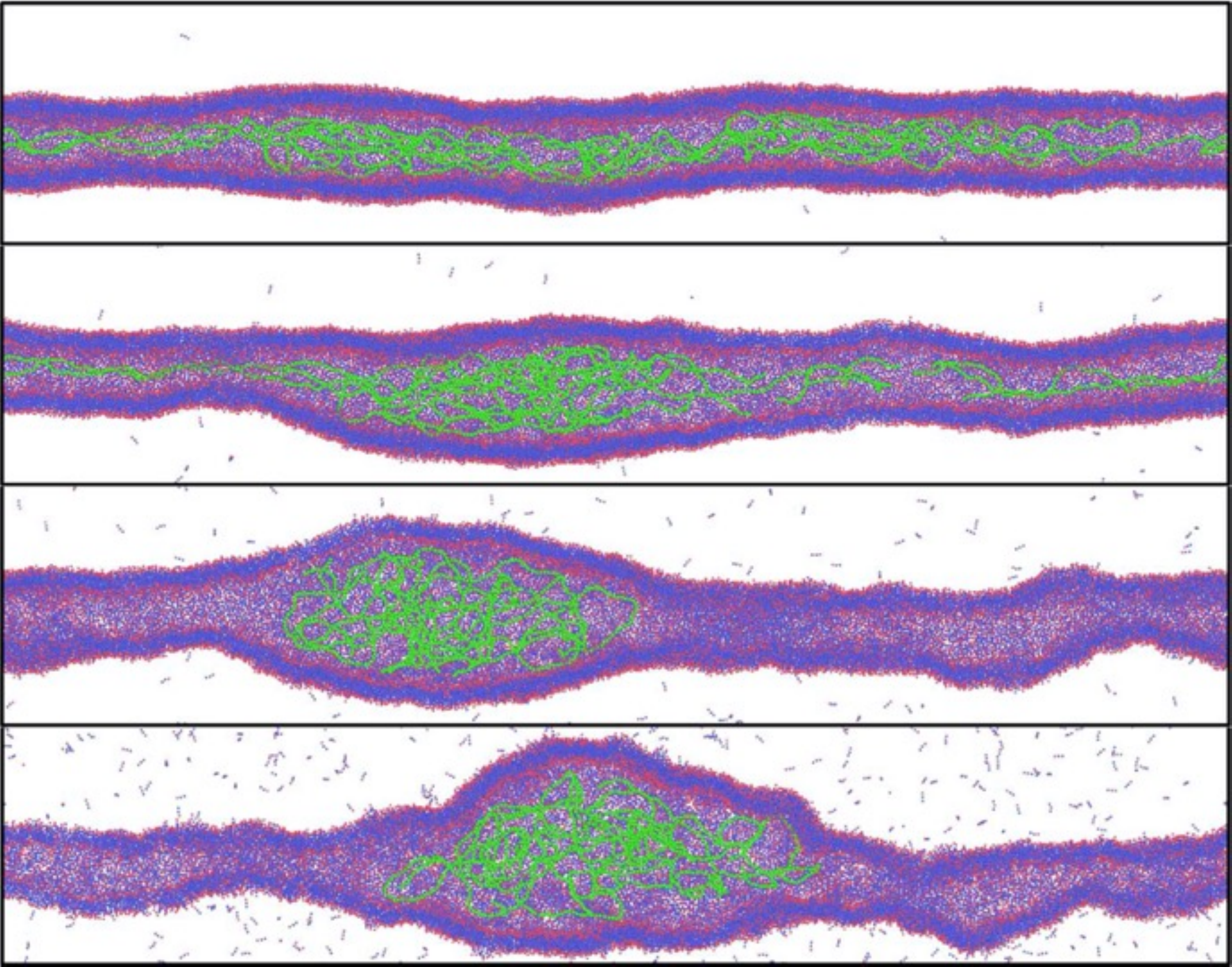}\hfill%
\includegraphics[width=.33\textwidth]{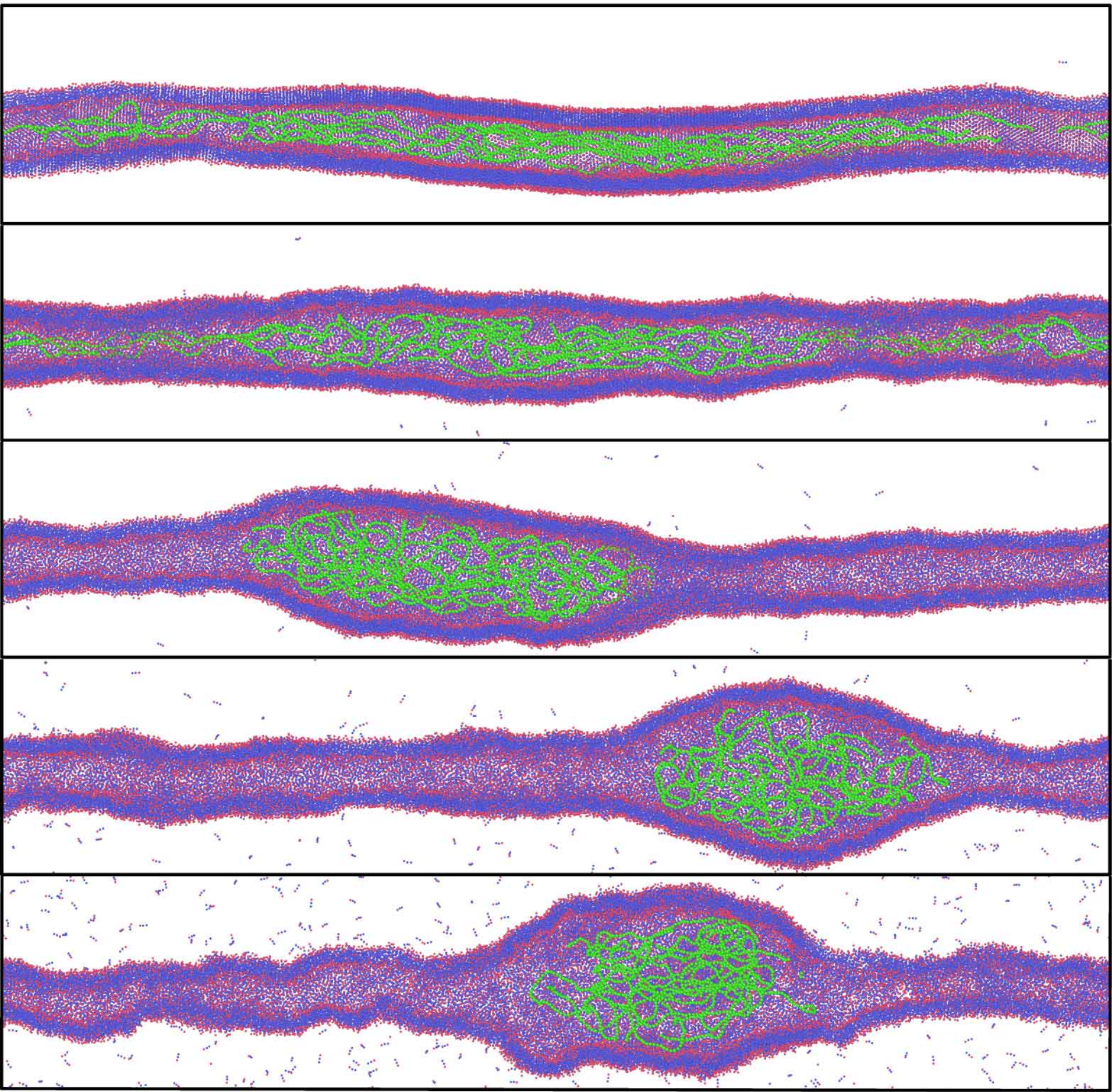}\\
(a) $w_c = 1.3 \sigma$~~~~~~~~~~~~~~~~~~~~~~~~~~(b) $w_c = 1.4 \sigma$~~~~~~~~~~~~~~~~~~~~~~~~~~(c) $w_c = 1.5 \sigma$
\vspace{0.1cm}
\caption{Simulation snapshots at $t = 5 \times 10^6 \Delta t$ of ``swollen'' and ``globular'' equilibrium conformations of a polymer chain in a bilayer tubule having different bending moduli $\kappa$ and tension $\Sigma$ obtained by varying the range of lipid interaction potential $w_c$ among tail beads and temperature $k_B T/\epsilon$. The bending modulus $\kappa$ of the tube is obtained by fitting the fluctuation spectrum to a functional form~\cite{p:cooke2005,p:fournier2007}.}\label{fig2}
\end{figure}

The \emph{coil} to \emph{globule} transition depends on the tubule properties too. In the limit of a rigid tube $\kappa >> k_{B} T$, polymer prefers to be in an extended state. In the membrane model, bending modulus of the membrane can be tuned by tuning the attractive interaction between the tail beads, \emph{i.e.} by changing potential width $w_c/\sigma$ and the effective temperature $k_{B} T/\epsilon$. We have obtained a phase diagram to explore \emph{coil} to \emph{globule} transition in the $w_c/\sigma$ -$kT/\epsilon$ plane. The final configuration snapshots after $t = 5\times 10^6$ time steps are shown in the Fig.~\eqref{fig2}. Corresponding temperatures $k_{B} T$, bending modulus of the tubule, $\kappa$ and the averaged tube radius squared $\langle R^{2}_{0} \rangle$ are shown in the Table \eqref{table_1}. Beyond the range of the temperature $k_{B} T$ as shown Table \eqref{table_1}, \emph{i.e.} for higher temperature bilayer tube is not stable.

\setlength{\arrayrulewidth}{0.3mm}
\setlength{\tabcolsep}{18pt}
\renewcommand{\arraystretch}{1.2}
\begin{table}[!h]
\centering
\begin{tabular}{ |p{3cm}|p{3cm}|p{3cm}|p{3cm}|  }
\hline
Tube properties &$w_c = 1.3 \sigma$ & $w_c = 1.4 \sigma$ & $w_c = 1.5 \sigma$\\
\hline
\emph{(i)} $k_B T /\epsilon$ &  &   & 0.9\\
$~~~~\kappa /k_B T$           & & & 28.7667 \\
$~~~~\langle R_0^2 \rangle / \sigma$            &  &  &79.8670 \\
\hline
\emph{(ii)} $k_B T /\epsilon$ & &  0.9 & 1.0\\
$~~~~~~\kappa /k_B T$           &  &  15.7465 & 18.4059 \\
$~~~~~~\langle R_0^2 \rangle / \sigma$            &  & 104.3290 & 102.4582 \\
\hline
\emph{(iii)} $k_B T /\epsilon$ & 0.8 &  1.0 & 1.1\\
$~~~~~~~\kappa /k_B T$           &  14.2495 &  14.1763& 16.4098 \\
$~~~~~~~\langle R_0^2 \rangle / \sigma$            & 106.9920 &119.2689 & 114.9923 \\
\hline
\emph{(iv)} $k_B T /\epsilon$ & 0.9 & 1.1 & 1.2 \\
$~~~~~~\kappa /k_B T$             & 11.9555 & 11.3364  & 12.3305 \\
$~~~~~~\langle R_0^2 \rangle / \sigma$  & 124.1086 & 133.5598  & 126.8585 \\
 \hline
\emph{(v)} $k_B T /\epsilon$ &  1.0 &  1.2 & 1.3 \\
$~~~~~\kappa /k_B T$             & 9.9009 &  5.6646 & 7.9215 \\
$~~~~~\langle R_0^2 \rangle / \sigma $  & 141.1741 & 144.8585 & 135.7489 \\
\hline
\end{tabular}
\caption{Bending rigidity $\kappa$, and square averaged tubular radius $\langle R_{0}^{2} \rangle$ for tubules shown in Fig.~\eqref{fig2} without an encapsulated polymer. The number of lipids used in simulating the tubules are $N_t = 97200$.}\label{table_1}
\end{table}

\subsection{Statistical properties of Tubule:}
\begin{figure}[!h]
\subfloat[]{\includegraphics[width = 0.33\textwidth]{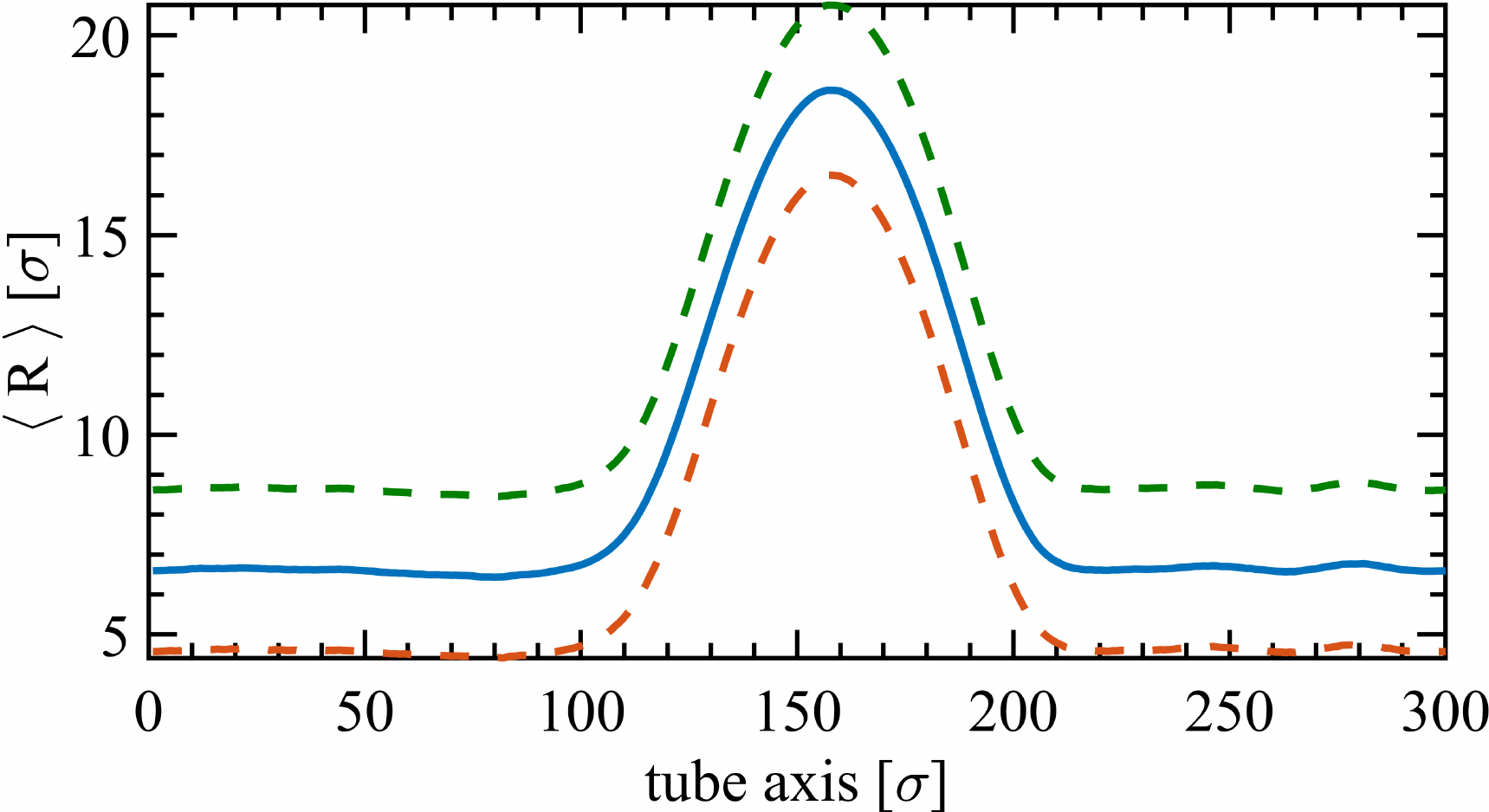}}
\subfloat[]{\includegraphics[width = 0.33\textwidth]{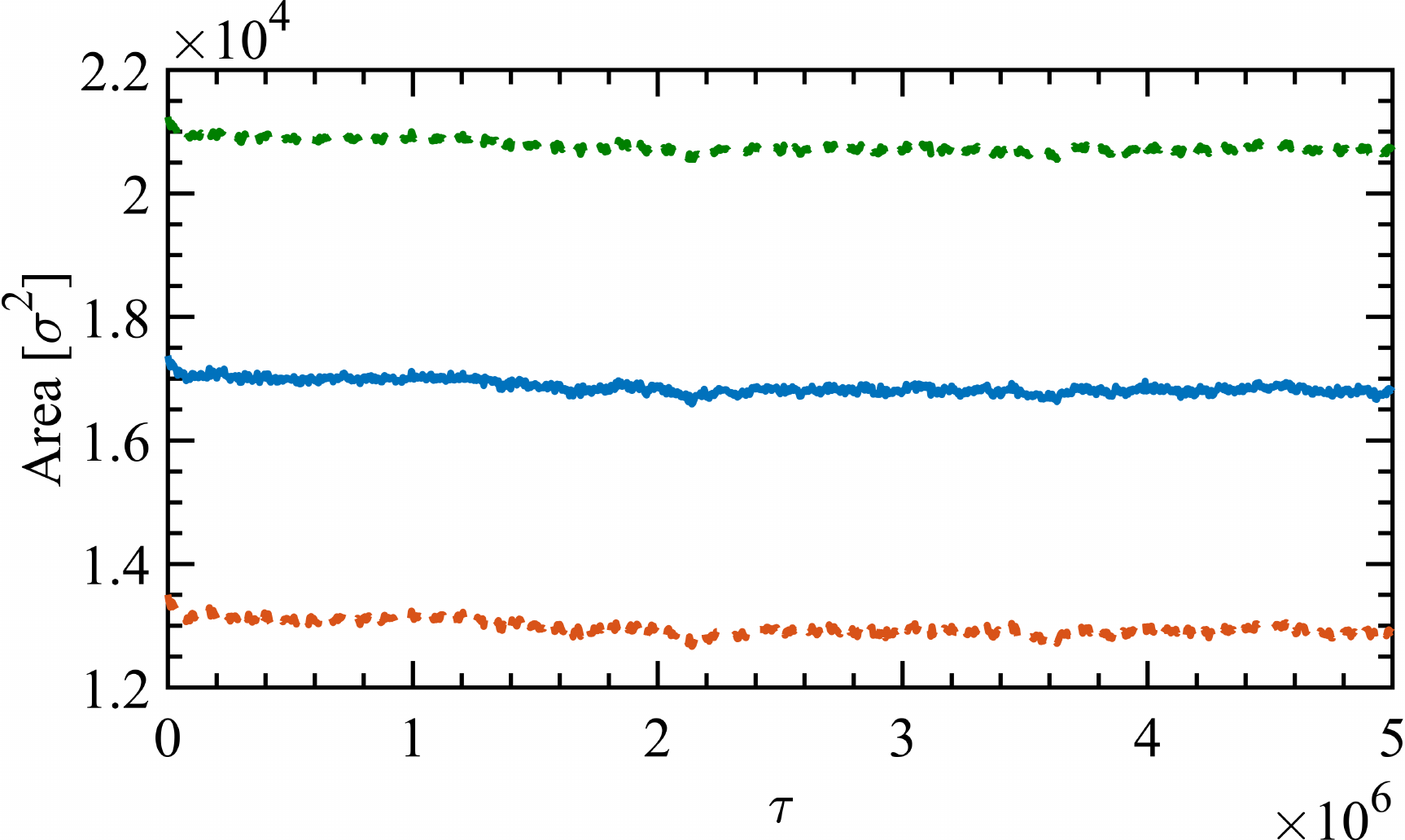}}
\subfloat[]{\includegraphics[width = 0.33\textwidth]{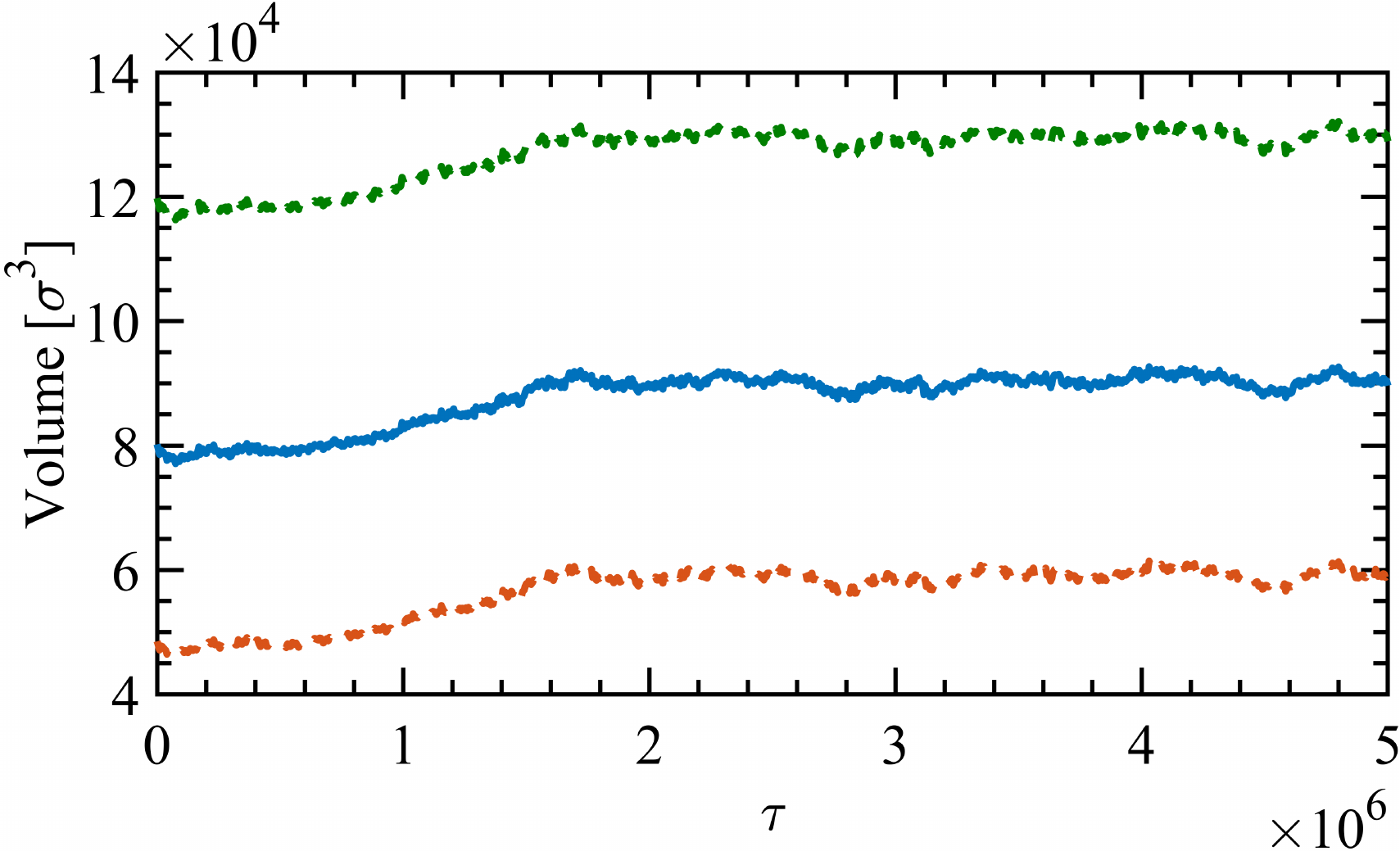}}
\caption{\emph{(a)} Radial bulge along the axis of the tube $x$ for outer $\langle D_{out}/2 \rangle$ (dashed orange), and inner $\langle D_{in}/2 \rangle$ leaflets (dashed purple), and the \emph{i.e.} average radius $\langle D(x)/2 \rangle$. Panel \emph{(b)} shows that the area of the bilayer tubule remains constant over time. Panel \emph{(c)} shows that the volume of the tube increases during the coil to globule transition.}\label{fig3}
\end{figure}

The bilayer tubule used in our simulations has a finite thickness, with the outer and inner diameters being $D_{out}$ and $D_{in}$ respectively. In order to determine the radial bulge of the tubule we divide the cylindrical tubule in slices of width $\sigma$, \emph{i.e.} size of a head bead. Therefore the tubule can be visualised of as a series of adjacent rings of width $\sigma$. The instantaneous tube diameter at position $x$ is calculated by taking the mean of the inner and outer $D(x) = \left( D_{in}(x) + D_{out}(x) \right)/2$ diameters. For a fluctuating tubule the $D_{out}$ and $D_{in}$ are the distance of the head beads from the centre of mass of each ring. We compute $\langle D(x) \rangle$ by performing a time average of the instantaneous tube diameter measured every $\tau_{m} = 10^3 \Delta t$ time steps after equilibration $\tau_{eq}$. For axisymmetric tubular shapes $\langle D_{in}(x)/2 \rangle$ and  $\langle D_{out}(x)/2 \rangle$ is independent of the azimuthal angle of a head bead on the ring. This is seen for the tubule without a confined polymer and for ellipsoidal tubular shapes. However for toroidal coils $l_p >>  R_{0}$ the azimuthal symmetry is lost. In this case we average over the azimuthal angle for all head beads in a given ring along the tubule, to obtain instantaneous values of the tubule diameter $D_(x)$ and in turn its thermal average $\langle D(x) \rangle$. Fig.~\eqref{fig3}(a) shows the time averaged radial bulge of an axisymmetric tubule along the tube axis having a confined polymer. The outer and inner inner leaflets $\langle D_{out}/2 \rangle$ and $\langle D_{in}/2 \rangle$ are shown by dashed lines. The area of the tubule remains constant throughout the simulation (Fig.\ref{fig3}(b)) while conformational transition of the chain is accompanied by a volume increase (Fig.\ref{fig3}(c)). 

\begin{figure}[h]
\centering
\includegraphics[width =0.7\textwidth]{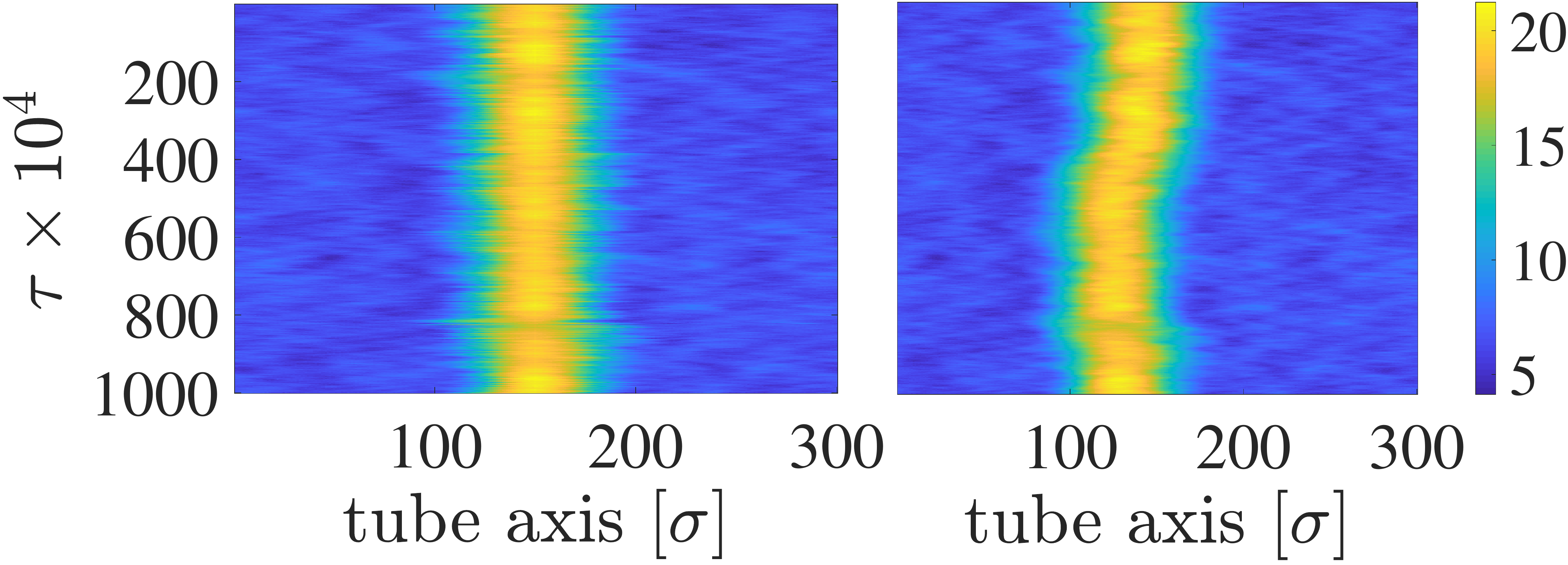}
\caption{Space time plots of the radial bulge of a deformed tube along the tube axis (right). Colormap indicates the value of the radius of each tubular cross section (ring of lipid molecules). In order to compute the time average the deformed profile at each time instant is shifted along the tube axis such that the maximum deformation (brightest point) occurs at $x=L/2$ (left).}
\label{fig04}
\end{figure}

The centre of mass of each ring of lipid molecules that form the tubule is allowed to fluctuate. Fig.~\eqref{fig04} (right panel) shows the radial deformation profile along the tube axis as a function of time. Due the motion of the centre of mass of the deformed segment of the tubule along the tube axis (SI movies, and Fig.~\ref{fig2}) the maximum radial bulge occurs at different positions $x$. A simple averaging of the deformed configurations over time to yield $\langle D_{m}/2 \rangle$ would therefore be erroneous. Therefore in order to obtain the accurate measure of the radial bulge (and since we have periodic boundary conditions imposed along the direction of the tube) we have shifted the deformed tubular profile, along the tube axis such that the maximum deformation occurs at $x=L/2$. 

\subsection{Statistical properties of Confined Chains: long length scales $l \sim \xi$}

The statistical properties of the polymer chain in the long wavelength limit $l \sim \xi$ \emph{e.g.} $\langle R_{\perp} \rangle$, the maximum extent of the polymer along radial direction and  $\langle R_{||} \rangle$ the maximum extent of the polymer along axial direction, are calculated by noting the position of the terminal monomers of the deformed chain. In order to compute $\langle R_{\perp} \rangle$ we have averaged over the radial coordiantes \emph{i.e.} $x$-$y$ plane. Monomer position snapshots are collected every $10^{3} \Delta t$ time steps once equilibrium $t = \tau_{eq}$ is reached. 

\subsection{Statistical Properties of Confined Chains: short distance $l_{p} \sim l \lesssim \xi$}

\subsubsection{Asphericity $\Delta$ and the nature of asphericity ($\Xi$)}
Polymer shapes are quantified by two parameters, asphericity ($\Delta$) and the nature of asphericity ($\Xi$)~\cite{p:alim2007}.These quantities are computed from the gyration tensor (see text). Intuitively, the parameter $\Delta$ is related to the variance of the three eigenvalues $\sum_i (\lambda_i-\bar{\lambda})^2$, where $\lambda_i = \lambda_x, \lambda_y, \lambda_z$ and $\bar{\lambda} = (\lambda_x + \lambda_y + \lambda_z)/3$, while nature of asphericity $\Xi$ is related to the product of the eigenvalues $\prod_{i}(\lambda_i-\bar{\lambda})$. To visually describe the shapes, we rescale these two parameters $2 \sqrt{\Delta}$ ranges $[0~2]$ and $\cos^{-1}\Xi/3$ ranges $[0~ \pi/3]$~\cite{p:alim2007}. The geometric interpretation of the shapes shown in the Fig.~\ref{fig:table_figures} (in the main text). The value $\Delta = 1$ is rod like shape and $\Delta = 0$ is fully spherical shape, whereas, $\Xi$ measures whether it is prolate or oblate. 

\subsubsection{Pair correlation function $g(r)$}
\begin{figure}[!h]
\subfloat[]{\includegraphics[width = 0.24\textwidth]{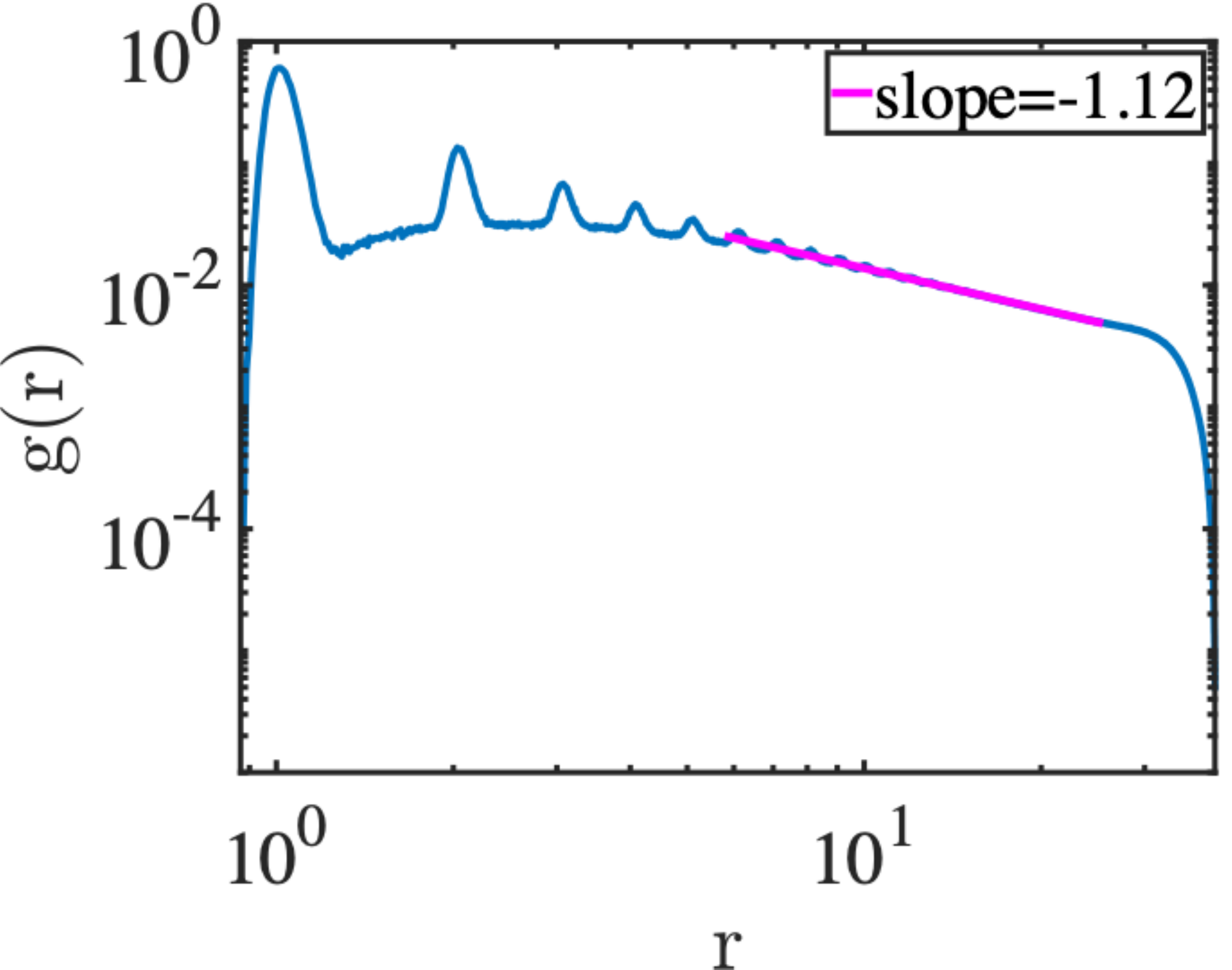}}
\subfloat[]{\includegraphics[width = 0.24\textwidth]{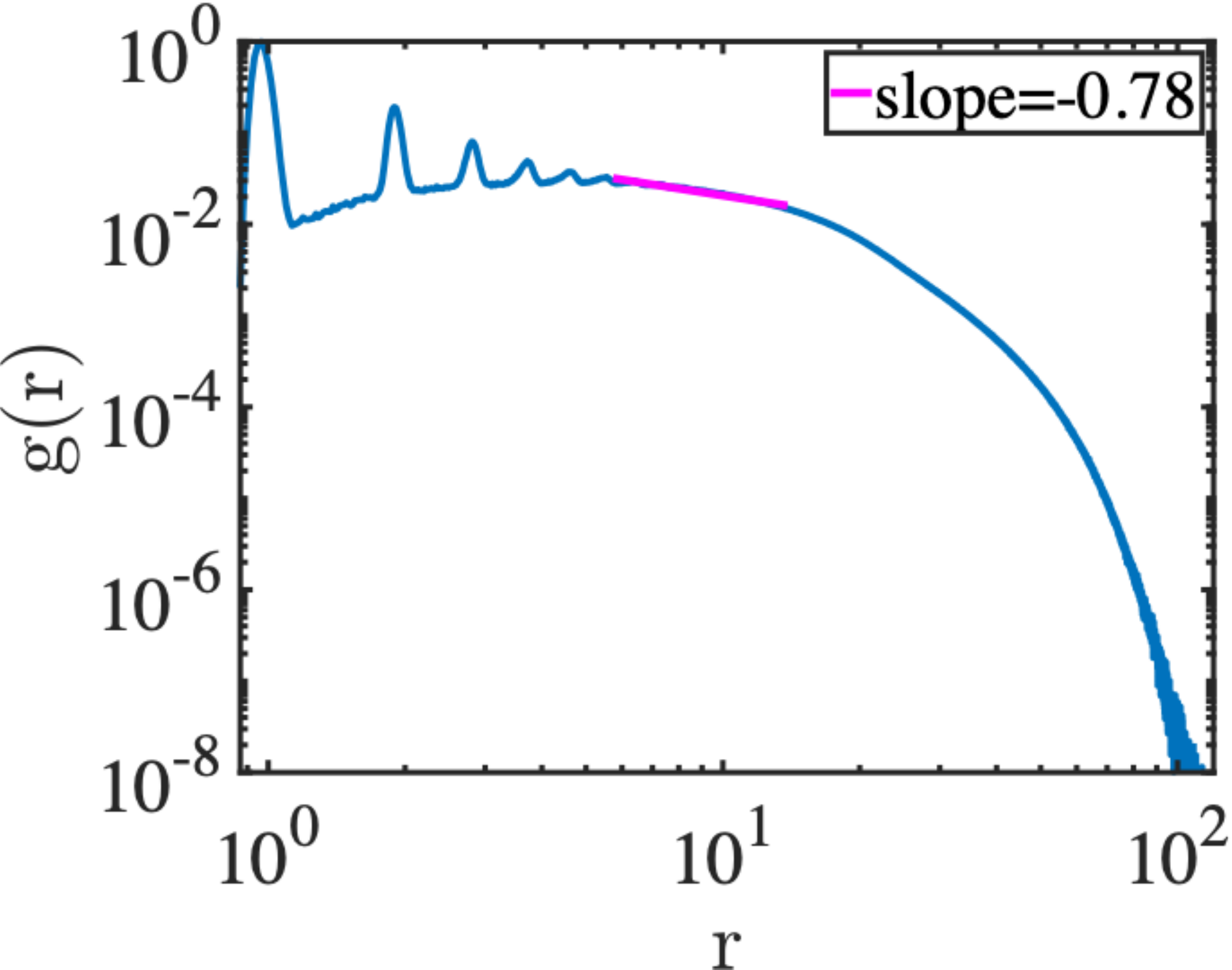}}
\subfloat[]{\includegraphics[width = 0.24\textwidth]{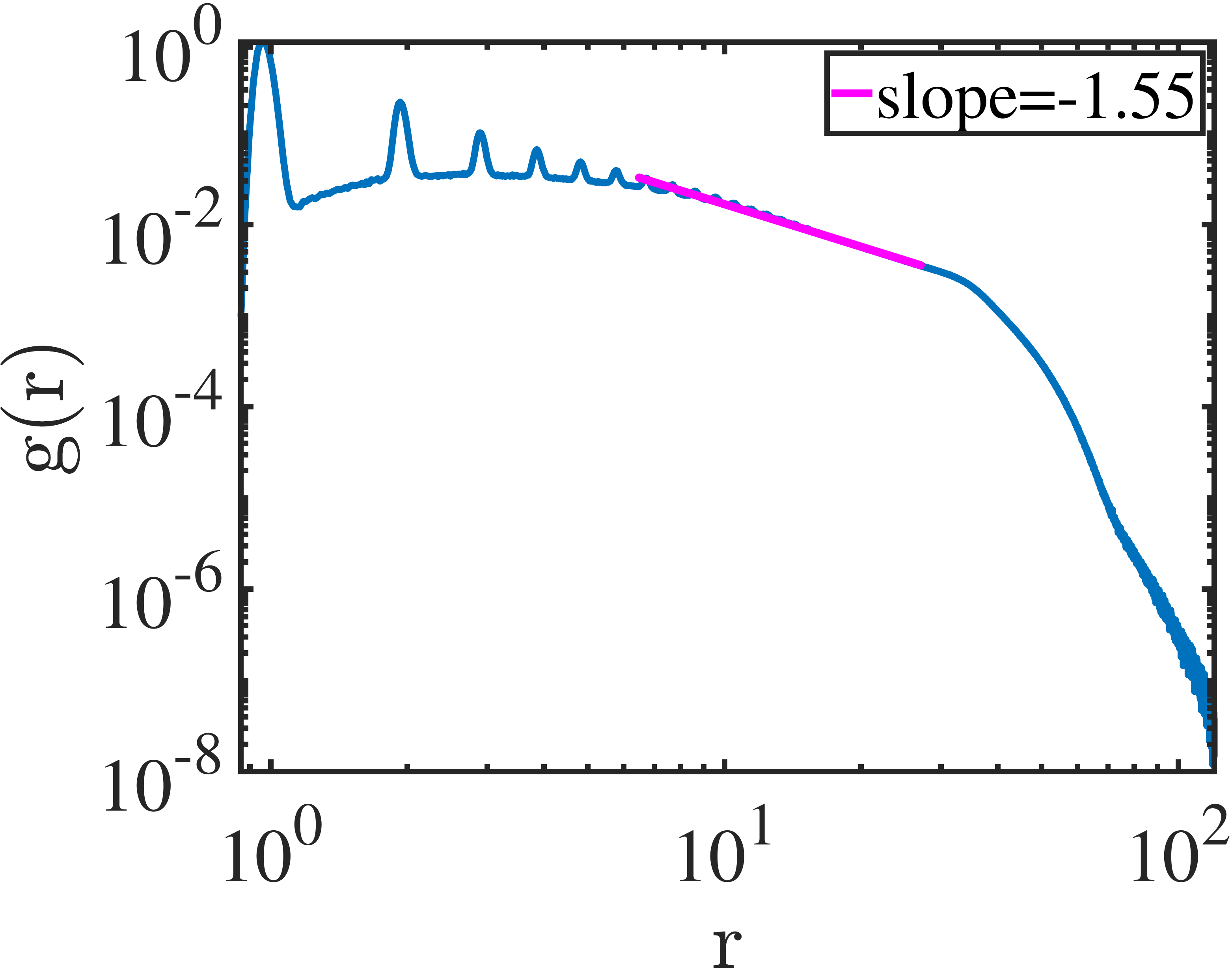}}
\subfloat[]{\includegraphics[width = 0.24\textwidth]{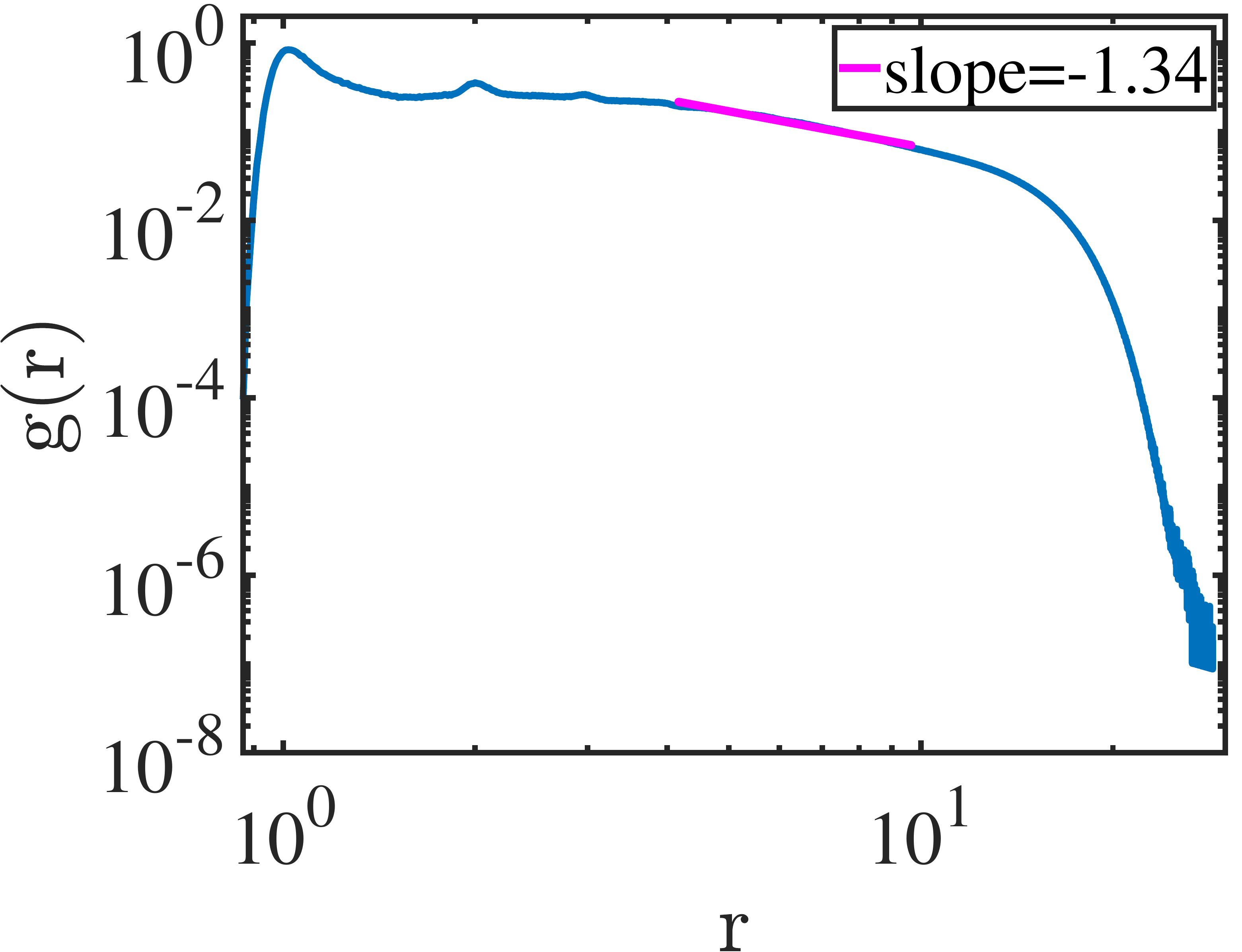}}
\caption{Pair correlation function for compact polymer conformations inside a soft tubule corresponding to panels \emph{(a), (e), (i)} and \emph{(m)} of Fig.~\eqref{fig:table_figures} respectively.}
\label{SI_03_gr}
\end{figure}

The pair correlation function or radial distribution function $g(r)$ describes how density of the monomers varies as a function of distance from a reference particle. We compute the number of particles in the interval between $r$ and $r+dr$ from a reference particle. Averaging over all monomers present in the system gives $g(r)$. In other words $g(r)$ is the ratio between the average number density at a distance $r$ from any given monomer as shown in the Fig.~\eqref{SI_03_gr}. 
The pair correlation function for different geometric constraints corresponding to Fig.~\ref{fig:table_figures} is shown in Fig.~\ref{SI_03_gr}. The positions of the peaks in $g(r)$ indicates the length scale associated with range of order in the system from which the structure factor $S(q)$ can be computed. Scattering experiments can then be used to interrogate the numerical data based on which different compact confined polymer structures can be explored. A full exploration of the static and dynamic structure factors, $S(q)$, and $S(q, \omega)$ is relegated to a future study~\cite{p:biswas2020}.

\subsubsection{Radial monomer density $\rho(r)$}

Density distribution of a polymer chain in different environments averaged over $\tau_m = 5000 \Delta t$ timesteps is shown in Fig~\ref{fig:table_figures} (main text). We compute the distance of a monomer $r$ from the centre of mass coordinate $r_i = \sqrt{(x_i-\bar{x})^2+(y_i-\bar{y})^2+(z_i-\bar{z})^2}$. A histogram obtained by binning the monomers as a function of $r$ gives density distribution $\rho(r)$. In case of strong confinement, the monomer density is high near the boundary, while for a weak confinement the monomers distribution is peaked about the centre of the channel. The monomer density is used to distinguish between ellipsoidal and toroidal coils (main text).

\section{Shape equation of tubules:}

\subsection{Tubular lipid membrane}

We develop a variational formulation to predict axisymmetric deformed tubule shapes (altered via membrane insertion) by minimising the membrane free energy~\cite{p:deuling1976}. A constant pressure $\Delta P$ acts across the membrane. Motivated by simulations, we work in an ensemble where the constant surface tension $\Sigma$ and pressure difference $\Delta P$ is constant. The elastic energy to deform a tubular membrane accounting for changes in area and volume, is given by the Helfrich-Canham free energy~\cite{p:deuling1976} 

\begin{equation}\label{eq:3.1}
\mathcal{F} = \frac{\kappa }{2}\int \left(c_1 +c_2 - c_0 \right)^2 dA + \Sigma \int dA - \Delta P \int dV - f L,
\end{equation}
where $c_1$ and $c_2$ are the the two principal curvatures of the axisymmetric tubule of length $L$ and $c_0$ is the spontaneous curvature. For a bilayer membrane $c_0 = 0$. An energetic cost of pulling out a tether of length $L$, from a GUV, relevant in micropipette aspiration experiments is also included. For a cylindrical shape $c_1 = \frac{1}{R}$ and $c_2 = 0$, thus equilibrium free energy of the cylindrical membrane tube can be written as, $\mathcal{F}_{\text{eq}} = \frac{\pi \kappa L}{R} +2 \pi R L \Sigma  - \pi R ^2 L \Delta P -f L$. For an unperturbed tube $\Delta P = 0$,  $\frac{\partial \mathcal{F}_{\text{eq}}}{\partial R}=0$ and $\frac{\partial \mathcal{F}_{\text{eq}}}{\partial L}=0$ gives equilibrium radius $R_0 = \sqrt{\frac{\kappa}{2\Sigma}}$ and tube axis directional pulling force $f_0 = 2 \pi\sqrt{2 \kappa~\Sigma}$~\cite{p:derenyi2002}. The radius $R_0$, and critical pulling force $f_0$ depends on pressure $\Delta P$ for membrane parameters remaining constant \emph{i.e.} $\kappa$ and $\Sigma$.

\begin{figure}[h]
\begin{center}
\includegraphics[width = 0.8\linewidth]{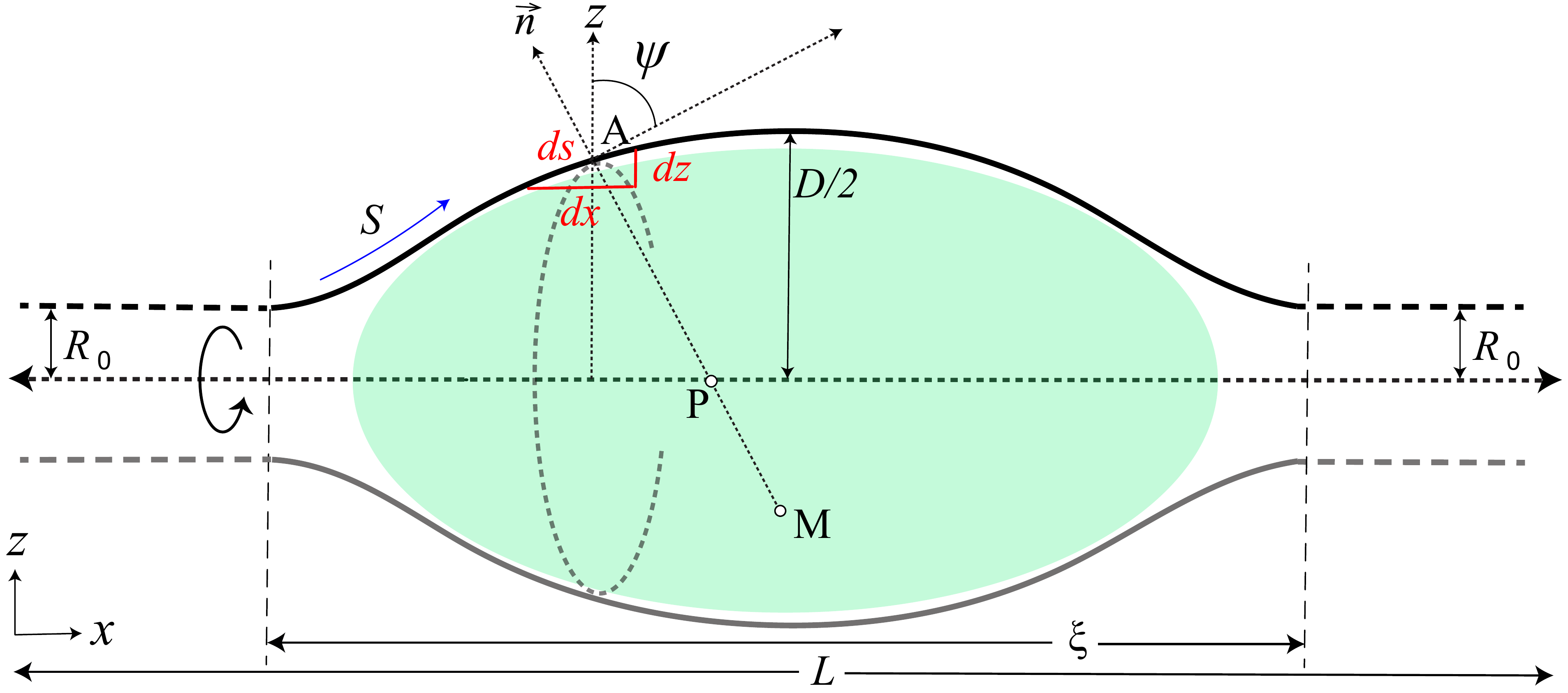} 
\caption{Parameterization of tubular shape. $ X $ is axial direction of the tube, the tube is assume to be symmetric about this axis. Tube forms bulge along $Z$ axis which is perpendicular to the $X$ axis. $\psi(S)$ is the angle between tangent at point A on the curve and the $Z$ axis.  $S$ is the arc-length of the contour.}
\label{fig:tubeschematic}
\end{center}
\end{figure}

Solutions to variational shape equations for axisymmetric tubules with open boundaries have not been explored in detail. These stem from a lack of understanding of the proper boundary conditions that need to be enforced to solve the first variation of the Helfrich-Canham free energy~ \cite{p:julicher1994, p:zhong1987}. A schematic figure showing an axisymmetric bilayer tubule deformed by an encapsulated polymer chain is shown in Fig.~\eqref{fig:tubeschematic}. We define a parameter $\delta = \frac{\xi}{L}$, where $\xi$ is the extent of deformation in the axial direction. This length scale can be compared with the aspect ratio of the tube $\rho = R_0/L$. The deformed tubular shapes depend on the ratio between these parameters $\delta/\rho$ (see Fig.~\ref{fig:tubeschematic}. Minimization of the curvature energy Eq.~\ref{eq:3.1} enforcing the boundary conditions and geometric and physical constraints (\emph{e.g.} constant area, volume) based on the experimental situation gives the equilibrium deformed axisymmetric tubule/vesicle shapes. 

\begin{figure}[h]
\begin{center}
\includegraphics[width = 95mm]{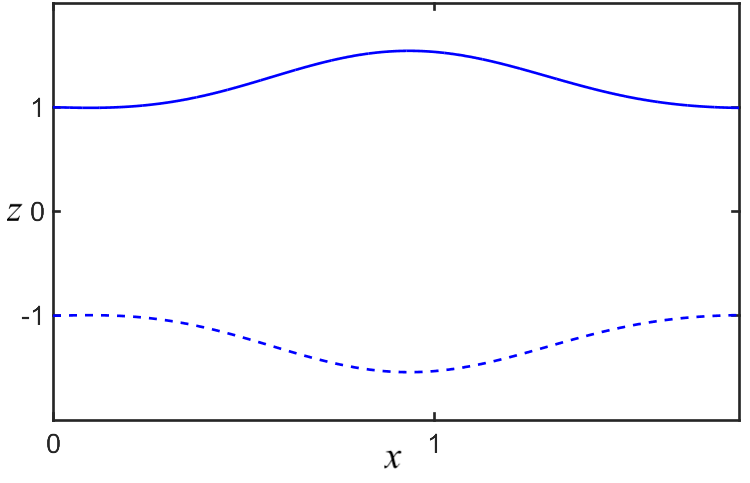}
\caption{Axisymmetric shape of the tubular membrane from the solution of the Eqs.~(\ref{eq:5.1.15}) with the initial conditions  $\psi =\frac{\pi}{2}, \dot{\psi} = 0, \tilde{\gamma } = 1.5, Z =1 $,$\tilde{P} = -10, \tilde{\Sigma} = 0.09$. We start the integration at $S = 0$ (as there are no divergent term for $Z \neq 0$) and stop the integration at $S = 3$ in which $\psi$ reaches at $\psi = \frac{ n\pi}{2}$. Dotted line is the reflection of the solid line about the mid-plane of the tubule.}
\label{fig:tubularshape}
\end{center}
\end{figure}

The curvature of $2$D surface embedded in $3$D space can be expressed in terms of two principal curvatures $c_1$ and $c_2$. Fig.~(\ref{fig:tubeschematic}) shows two principal radii of curvature $c_2 = \frac{1}{R_1}$ and $c_1 = \frac{1}{R_2}$ at point $A$. At point A, radius $R_1$ = AM corresponds to the radius of curvature along the meridian curve (M being the centre of the curvature of meridian curve on the plane of the figure). The radius of curvature $R_2$ = AP along the perpendicular direction of the figure has its centre at point P which lies on the axis of rotational symmetry along $X$ direction. From geometric representation of the curves we can write $dS = - R_1 d\psi $ and $Z = R_2 \sin \psi$. Thus, from the Fig.~(\ref{fig:tubularshape}), we can write down the geometric relations $\dfrac{\hbox{d}Z}{\hbox{d}S} = \cos \psi, \dfrac{\hbox{d}X}{\hbox{d}S} = -\sin \psi, c_2 = \dfrac{\hbox{d}\psi}{\hbox{d}S},$ and $c_1 = \dfrac{\sin \psi }{Z}$. We express the elastic free energy of the membrane tube in Eq.~\eqref{eq:3.1} using the above parameterisation, $\mathcal{F} =  2 \pi \kappa \int_{S_1}^{S_2} \mathcal{L}(\psi, \dot{\psi},Z,\dot{Z},\gamma)~dS$, where the \emph{Lagrangian} function~\cite{p:seifert1991}
\begin{equation}
\begin{split}
\mathcal{L}(\psi, \dot{\psi},Z,\dot{Z},\gamma) =  \frac{Z}{2} \left(\dot{\psi} + \dfrac{\sin \psi}{Z}  - c_0\right)^2 +\tilde{\Sigma} Z + \tilde{P} \frac{Z^2}{2} \sin \psi +\tilde{\gamma}(\dot{Z} - \cos \psi).
\end{split}
\end{equation}

The first variation of the free energy in Eq.~\ref{eq:3.1} yields one second-order (expressed in terms of two nested first-order differential equations) and two first-order differential equations in terms of the variables $\Psi(S)$, $U(S)$, $\tilde{\gamma}(S)$, and $Z(s)$~\cite{p:seifert1991} given by

\begin{subequations}\label{eq:5.1.15}
\begin{equation}\label{eq:5.1.15a}
\frac{\hbox{d} \psi(S)}{\hbox{d} S} =U(S), 
\end{equation}   
\begin{equation}\label{eq:5.1.15b}
\begin{split}
\frac{\hbox{d} U(S)}{\hbox{d} S} =- \frac{\cos{\left(\psi{\left(S \right)} \right)} U(S)}{Z{\left(S \right)}} + \frac{\sin{\left(2\psi{\left(S \right)} \right)} }{2Z^{2}{\left(S \right)}} + \dfrac{\tilde{\gamma }(S)\sin \psi(S)}{Z(S)} +\frac{\tilde{P} Z{\left(S \right)} \cos{\left(\psi{\left(S \right)} \right)}}{2},
\end{split}
\end{equation}  
\begin{equation}\label{eq:5.1.15c}
\frac{\hbox{d} \tilde{\gamma} (S)}{\hbox{d} S} =\tilde{P} Z{\left(S \right)} \sin{\left(\psi{\left(S \right)} \right)} + \tilde{\Sigma} + \frac{\left(U(S) - C_{0} \right)^{2}}{2} - \frac{\sin^2{\left(\psi{\left(S \right)} \right)}}{2 Z^2{\left(S \right)}},
\end{equation}
\begin{equation}\label{eq:5.1.15d}
\frac{\hbox{d} Z(S)}{\hbox{d} S} = \cos (\psi(S)).
\end{equation}    
\end{subequations}

We solve the coupled differential equations Eq.~\ref{eq:5.1.15} with initial conditions $\psi = \frac{\pi}{2}$, and $\dot{\psi} = 0$. In our numerical scheme we integrate Eq.~\ref{eq:5.1.15} starting from $S=0$ corresponding to $\Psi(s) = 0$, till $S \sim 1$, at which $\Psi = \pi/2$. This corresponds to a point $X=X_m$ along the tube axis. The profile is then reflected about a line perpendicular to the $X$-axis to generate the membrane configuration along the arclength $Z(s)$ above the midplane of the tube axes as shown in Fig.~\ref{fig:tubularshape}. A dotted line showing the reflection of the membrane profile obtained by reflecting $Z(S)$ about the midplane of the tubule is also shown. Taken together this represents the cross-section of the axisymmetric deformed tubule. We wish to compare equilibrium axisymmetric tubular shapes obtained from a minimisation of the curvature energy against those obtained from coarse-grained molecular dynamics simulations for specified value of surface tension and pressure difference across the membrane. An ideal scheme for predicting the tubular shapes observed in simulations with the variational formulation is to coarse-grain the tubular shapes, and fit it to the numerical solutions of the shape equations (with surface tension $\Sigma$ held fixed) to obtain the pressure difference $\Delta P$ required to bring about the shape changes. This pressure difference must exactly balance the entropic pressure exerted on the membrane walls by the polymer. However, due to convergence and stability issues of the numerical scheme used to solve the nonlinear shape equations (Eq.~\ref{eq:5.1.15}) for arbitrary values of pressure difference $\Delta P$, and surface tension $\Sigma$ we have not pursued this line of work in our study. 

\subsection{Pressure calculation from polymer statistics:} 
$F(N) = - k_{B} T \ln P(R)$, where $R = \sqrt{\langle R^{2} \rangle}$, mean end-to-end distance of the confined polymer chain. The volume of the ellipoid can be obtained a function of chain length $N$, $V(N)$. This can be used to compute the pressure exerted by the chain $P = - \frac{\partial F}{\partial V} = \frac{\partial F}{\partial N} / \frac{\partial V}{\partial N}$. Due to limitations in our system size $N$ and long simulation run times required to reach equilibrium we have relegated this to a future study. 

\section{Energy estimation of different shapes of the tubule}
We present a variational solution of the shape equation using two types of ansatz \emph{(a)} a spherical bulge as shown in Fig.~\ref{fig:bulge_geometry_1}, and \emph{(b)} a bell curve shaped bulge, on a cylindrical tube as shown in Fig.~\ref{fig:bulge_geometry_2}. Fig.~\ref{fig:bulge_geometry_2} resembles the geometry of a drop wetting a fibre~\cite{b:degennes2013}. The spherical bulge is parameterized by the radius $R$ of the sphere, and the diameter $D$, and length $L$ of the cylinder. Similarly the ``drop on a fibre'' geometry is parameterized by the axial scale of the bulge $\xi$, the maximal radial deformation $R+\frac{D}{2}$, and the length of the cylinder $L$.

\subsection{Spherical bulge model}
\begin{figure}[!h]
\centering
  \subfloat[]{\includegraphics[width=0.4\linewidth]{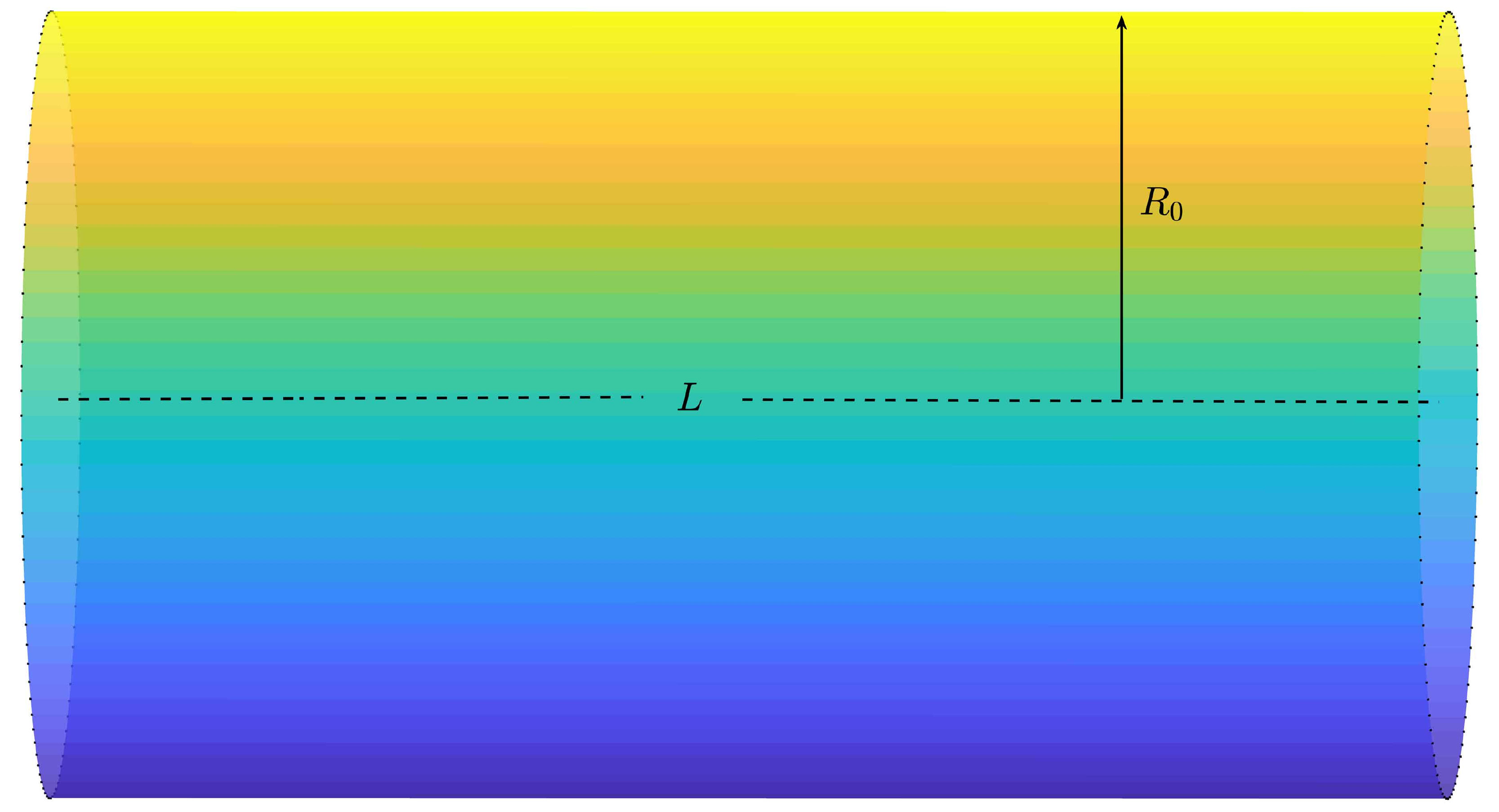}}
  \subfloat[]{\includegraphics[width=0.5\linewidth]{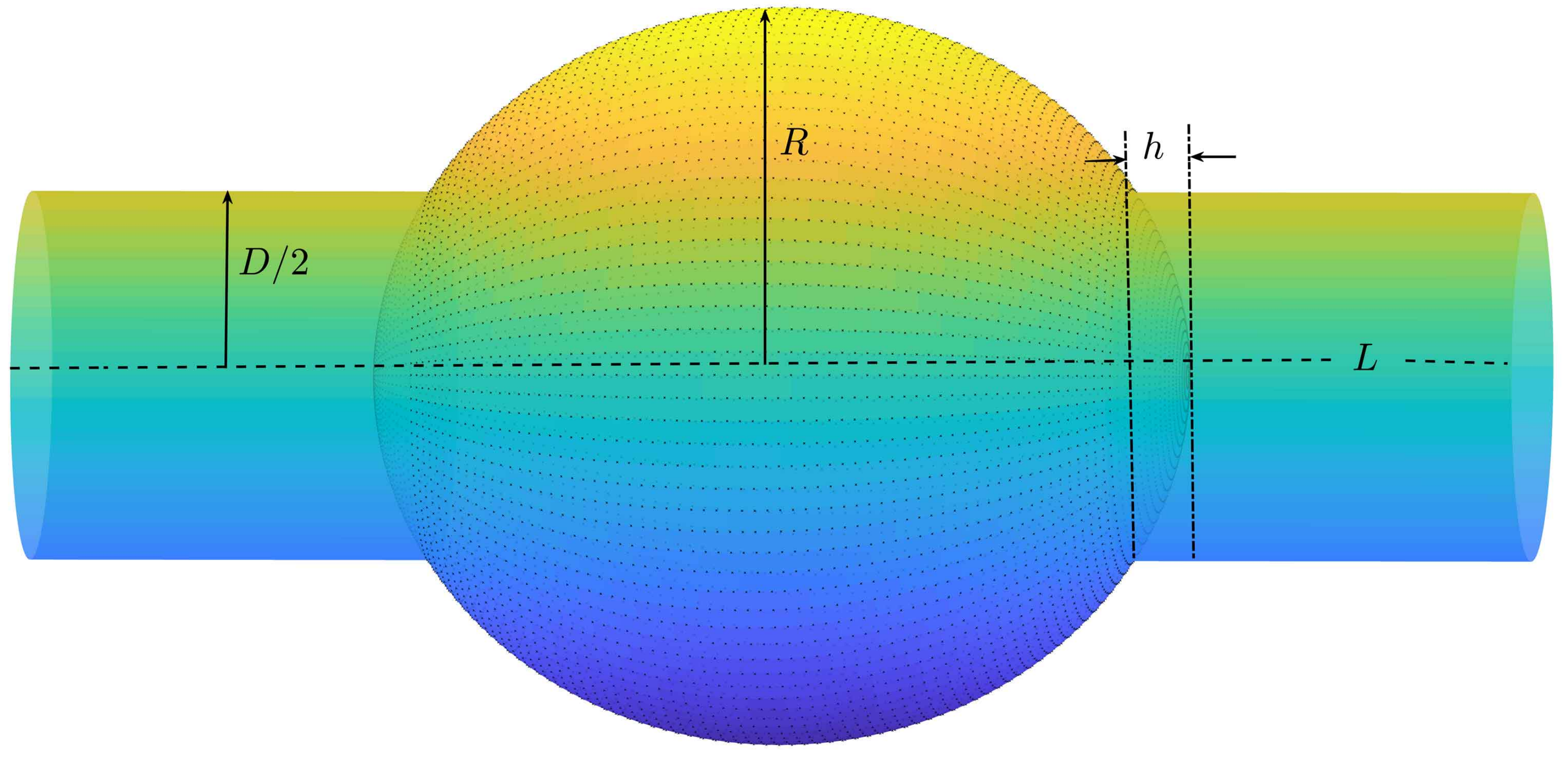}}
  \label{fig:sub2}
\caption{Schematic picture of (a) an undeformed cylindrical tube and (b) a cylindrical section of diameter $D$ with a deformed spherical bulge of radius $R$. Both cylinders have the same length $L$.}
\label{fig:bulge_geometry_1}
\end{figure}

Fig.~\ref{fig:bulge_geometry_1} shows a deformed tubule with a spherical bulge of radius $R$ in its centre; the deformation arising out of polymer insertion in the tubule. For the total area of the tubule to be conserved, the undeformed cylindrical section shrinks radially such that its radius $D/2 < R_0 $. Area of the undeformed tube is $2 \pi R_0 L$. Whereas area of the deformed tube is `area of the cylindrical part of length $(L - 2 R + 2 h)$ + area of the spherical part - $2 \times$ area of the spherical cap of radius $h$'.

Expressing $h$ in terms of $R$ and $D$, $h = R - \sqrt{R^2 - D^2/4}$, and noting that the area of the cylindrical tubular segment is $2 \pi R h$ as shown in Fig.~\ref{fig:bulge_geometry_1} the total area of the deformed tube with spherical bulge is given by

\begin{eqnarray}\label{eqn:1.1}\nonumber
   A_{sph} &=&  \pi D(L-2R+2h) + 4 \pi R^2 - 4\pi R h,\\\nonumber
           &=&  \pi D(L-2 \sqrt{R^2 - D^2/4}) + 4 \pi R^2 - 4\pi R(R - \sqrt{R^2 - D^2/4}),\\\nonumber
           &=&  \pi D L + 4 \pi \sqrt{R^2 - D^2/4}~(R - D/2).\\       
\end{eqnarray}

Next, we compute the energetic cost of deforming a cylindrical tubule to form a spherical bulge. Elastic energy of the tubule is given by, 
\begin{eqnarray}\label{eqn:1.2}
  F_b  = \frac{\kappa}{2} \int_{A} (c_1 +c_2)^2 dA,
\end{eqnarray}
where $c_1$ and $c_2$ are the principal curvatures and $\kappa$ is the bending modulus. For sphere of radius $R$, principal curvatures are $c_1 = c_2 = \frac{1}{R}$, while for a cylinder of radius $D/2$, the principal curvatures are $c_1 = \frac{1}{R}$ and $c_2 = 0$. Thus bending energy of the deformed tube is thus

\begin{eqnarray}\label{eqn:1.3}\nonumber
  F_b  &=& \frac{\kappa}{2} \left[ \frac{4}{D^2}\pi D(L-2 \sqrt{R^2 - D^2/4}) + \frac{4}{R^2} 4 \pi R^2 - \frac{4}{R^2} 4\pi R(R - \sqrt{R^2 - D^2/4})\right],\\\nonumber
  &=& 8 \pi \kappa \left[ \frac{(L-2 \sqrt{R^2 - D^2/4})}{4 D} +1 - \frac{R 
  -\sqrt{R^2 - D^2/4}}{R}\right],\\\nonumber
  &=& 8 \pi \kappa \left[ \frac{L}{4 D} + \sqrt{R^2 - D^2/4}\left(\frac{1}{R}-\frac{1}{2D} \right)\right],\\\nonumber
  \frac{F_b}{8 \pi \kappa}&=&  \frac{L}{4 D} + \sqrt{R^2 - D^2/4}\left(\frac{1}{R}-\frac{1}{2D} \right).\\
\end{eqnarray}

The elastic energy of the undeformed cylindrical tube computed from the Helfrich-Canham free energy Eq.~\ref{eq:3.1} is given by 
\begin{eqnarray}\label{eqn:1.4}
  \frac{F_{b}}{8 \pi \kappa}= \frac{L}{8R_0}.
\end{eqnarray}

\begin{figure}[h]
\centering
  \subfloat[]{\includegraphics[width=0.485\linewidth]{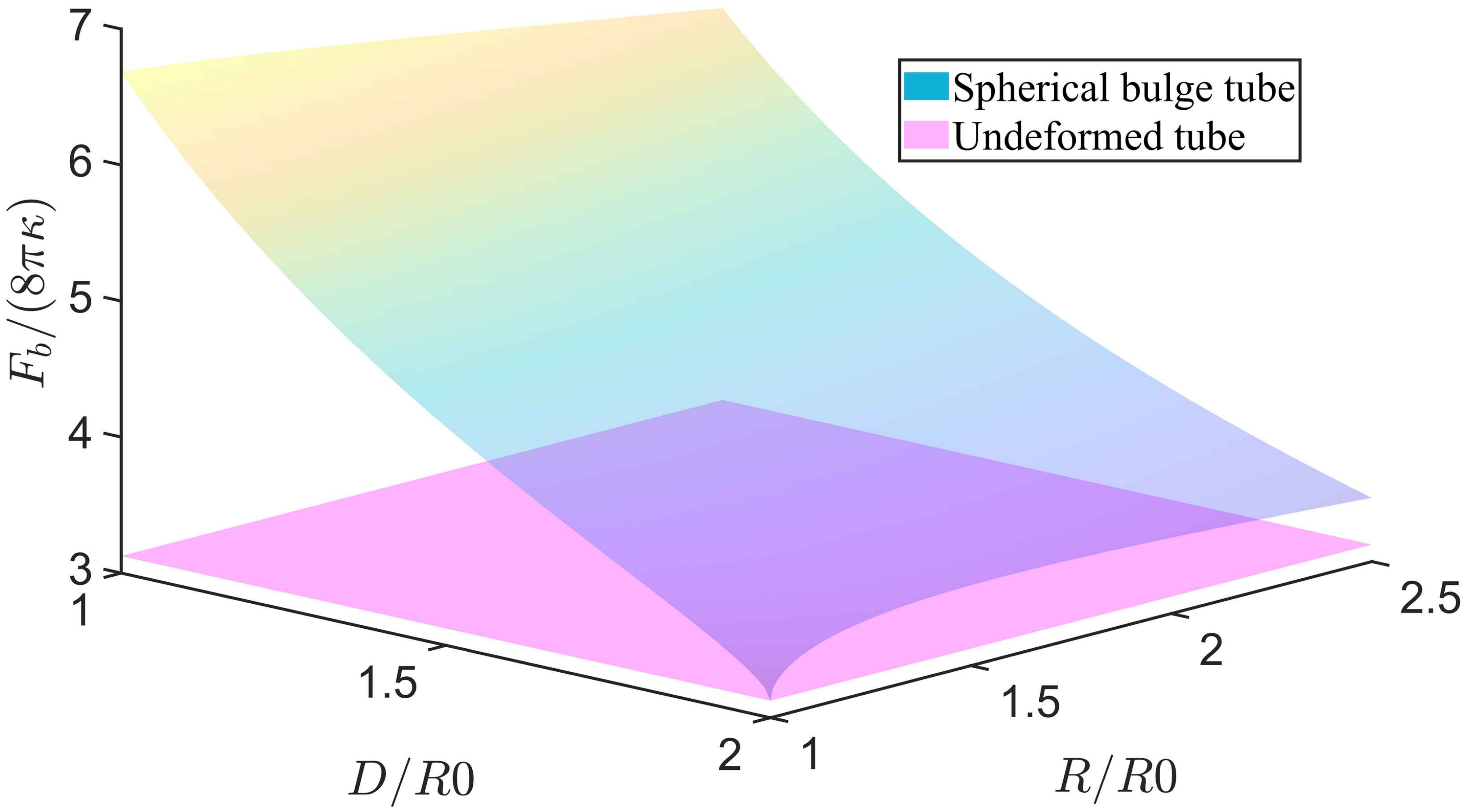}\label{fig:bulge_energy}}
  \subfloat[]{\includegraphics[width=0.5\linewidth]{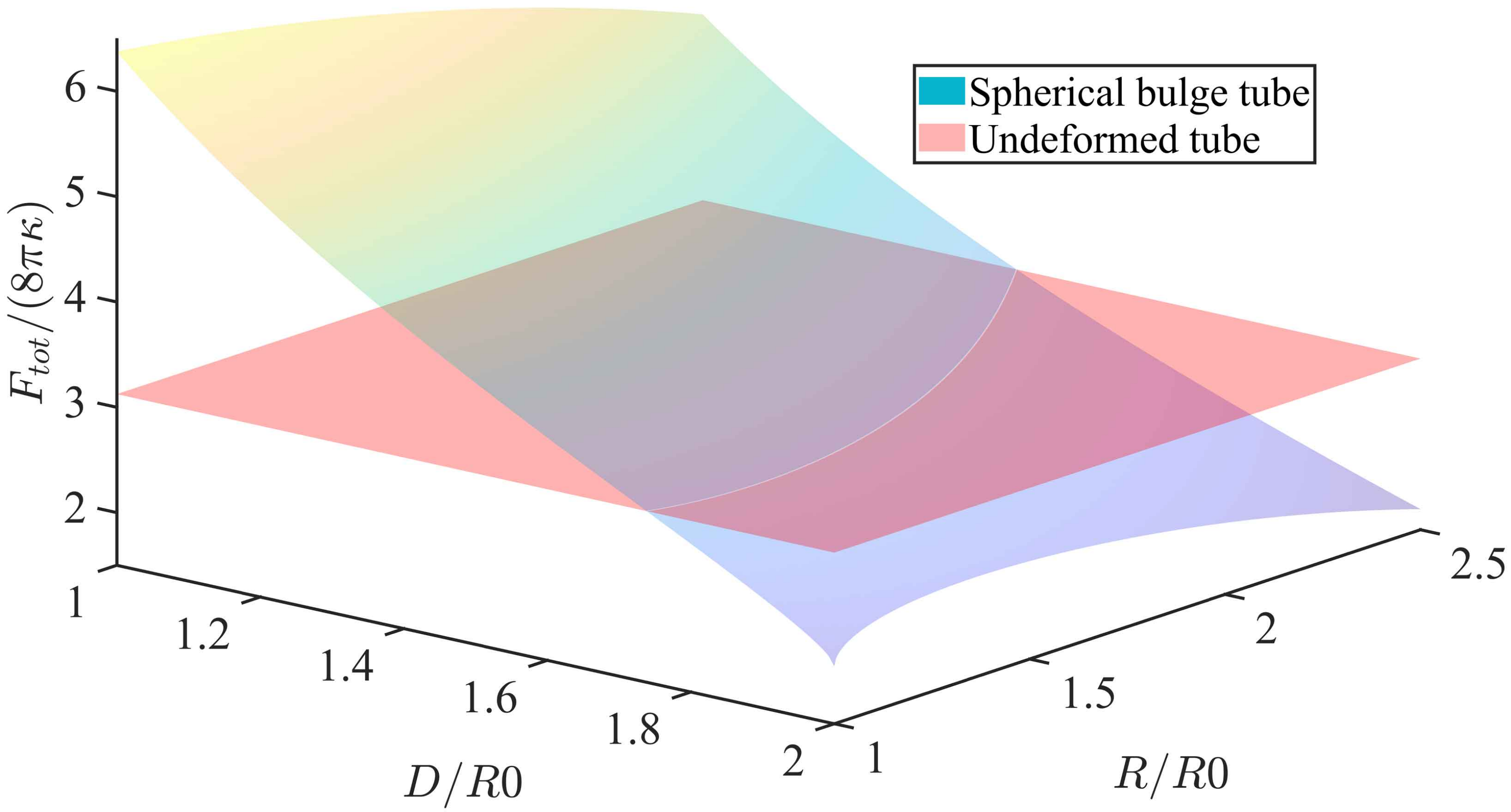}\label{fig:total_energy}}
\caption{ Surface plots of the elastic energy of (a) an undeformed cylindrical tube, compared against a deformed tube with a spherical bulge. The deformed tubule has a higher bending energy compared to the undeformed cylinder, with the energies being equal when $D/R_0 = 2$ and $R/R_0 = 1$. (b) Total energy of the deformed cylinder acounting for the work done by the increase in internal pressure due to the insertion of the polymer, for $\Tilde{p} = -0.0002$ compared against the energy of an undeformed cylindrical tubule. Two isosurfaces intersect where the total energy of the deformed tube is equal to the energy of the undeformed tube. The locus of intersection points traces a family of curves for which the energy of the deformed tubule is equal to that of the undeformed one.}
\end{figure}

The elastic free energy of the deformed cylinder Eq.~\ref{eqn:1.3} approaches that of the undeformed cylinder in the limit, $D/2 \rightarrow R_0$ and $R \rightarrow R_0$. This can be seen in Fig.~(\ref{fig:bulge_energy}). Further the elastic energy of the deformed cylindrical tubule is always greater than the undeformed one as shown in  in Fig.~(\ref{fig:bulge_energy}). In equilibrium, the pressure difference $\Delta p$ across the tubule is zero. Insertion of a polymer chain inside a tubule leads to an excess entropic pressure that brings about a change in volume. For a spherical bulged configuration to be energetically stable, the sum of the work done in deforming the tubule $\Delta p dV$ and the elastic energy cost of the deformed tubule must be equal to that of the undeformed cylinder. We thus scan the parametric space of two dimensionless parameters $D/R_0$ and $R/R_0$ numerically to find optimal shapes for the above condition to be satisfied. 

The work done by the excess pressure $\Delta p$ on the tubule is given by, 
\begin{eqnarray}\label{eqn:1.5}
  F_p  = - \Delta p \int_{v} dV.
\end{eqnarray}
For the spherical bulge geometry shown in the Fig.~(\ref{fig:bulge_geometry_1}) this yields,
\begin{eqnarray}\label{eqn:1.6}\nonumber
F_p &=& - \Delta p \left[ \pi \left( \frac{D}{2}\right)^2(L-2R+2h) + \frac{4}{3} \pi R^3 - \frac{2}{3} \pi h^2 (3R-h)\right],\\\nonumber
\frac{F_p}{8 \pi \kappa} &=& - \Tilde{p} \left[\frac{D^2 L}{32} + \frac{1}{6}\left( R^2 - \frac{D^2}{4}\right)^{\frac{3}{2}}\right],\\
\end{eqnarray}
where $\Tilde{p} = \Delta p / \kappa$. Thus total energy of the deformed tube is sum of bending energy and the work done by the excess pressure, $F_{tot} = F_b + F_p$. We plot $F_{tot}$ as a function of the dimensionless parameters $D/R_0$, and $R/R_0$. As seen in Fig.~(\ref{fig:total_energy}) $F_{tot}$ intersects with lowest energy state of the undeformed tube for a range of pressure values $\Tilde{p}$ values. Fig.~(\ref{fig:total_energy}) shows such an intersection point for $\Tilde{p} = 0.0002$.

\begin{figure}[h]
\centering
  \subfloat[]{\includegraphics[width=0.55\linewidth]{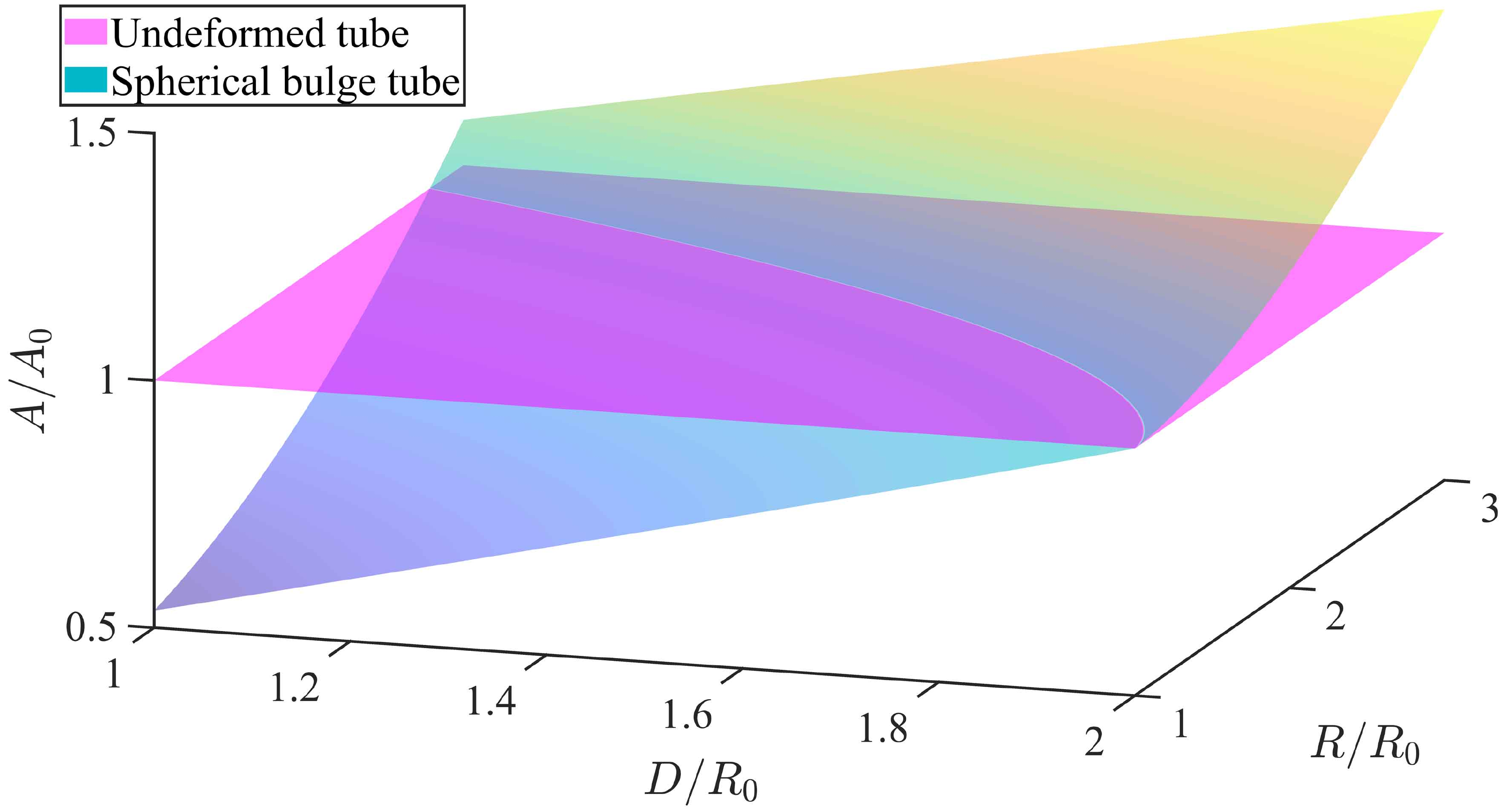}}
  \subfloat[]{\includegraphics[width=0.44\linewidth]{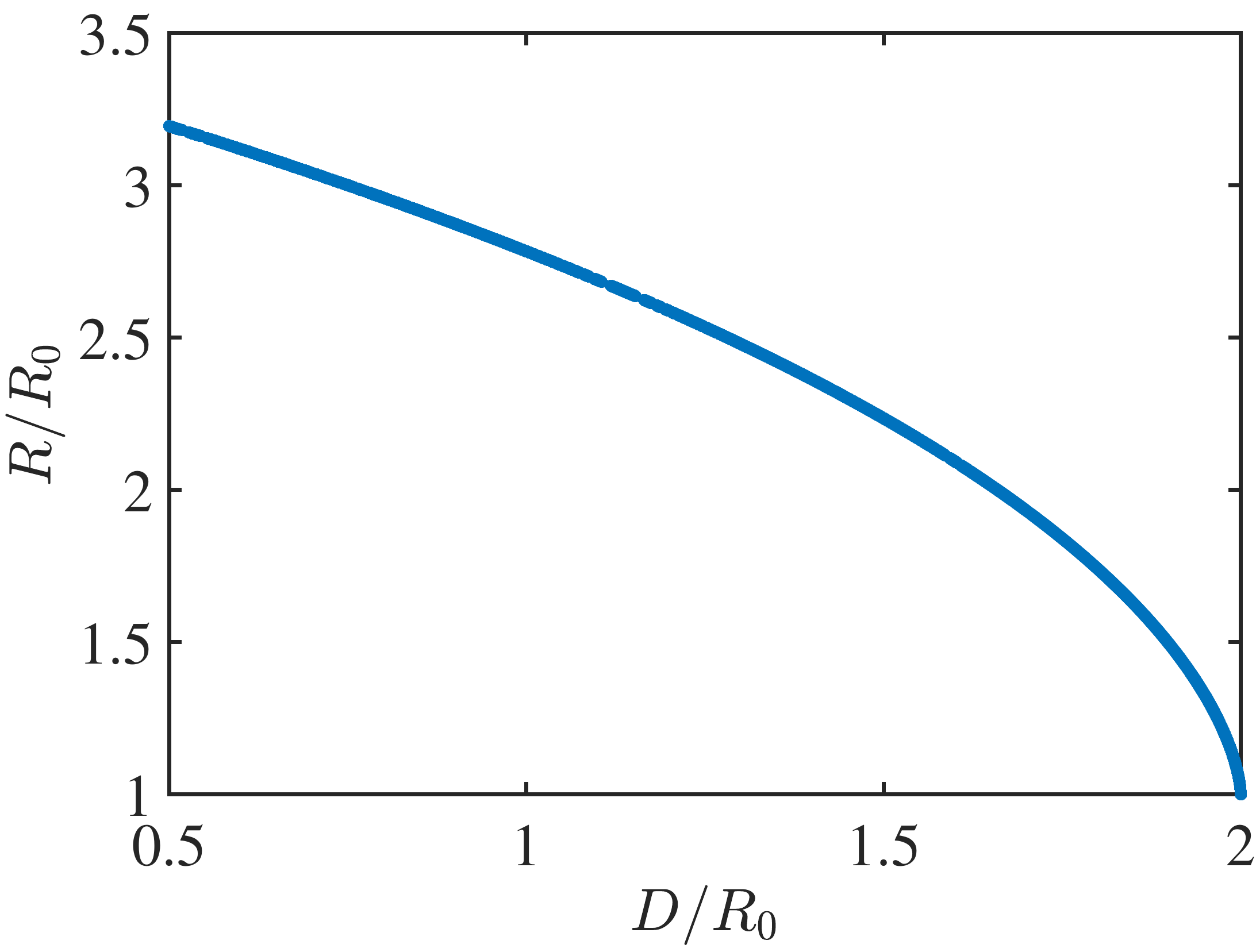} \label{fig:sub2}}
\caption{(a) Surface plots of the area of spherical bulge and undeformed cylindrical tube in the $D/R_0$-$R/R_0$ plane. The two surfaces intersect at points where the total surface area is equal to the area of the undeformed tube. (b) Locus of the intersection points of the two planes traces a family of curves in the parametric space.}
\label{fig:sphere_area}
\end{figure}

The intersection between the two planes traces out a curve yielding a family of shapes that minimise the Helfrich-Canham free energy. However, not all solutions satisfy the area constraint. The area of a deformed tubule can also be plotted in the $D/R_0$-$R/R_0$ parametric space following Eq.~(\ref{eqn:1.1}). The intersection between this area function and that of a cylindrical tubule traces out yet another family of curves for which the area of the deformed tube is equal to that of the undeformed cylinder as shown in Fig.~(\ref{fig:sphere_area}). Corresponding $D$ and $R$ values on the curve line maintain constant area constraint. Implementing constant area constraint in conjunction with the minimum free energy condition we obtain a deformed tubular profile for an imposed pressure $\tilde{p}$ for which the free energy is a minimum. The shape parameters $D/R_0$ and $R/R_0$ are uniquely determined for a given pressure $\tilde{p}$. 

\begin{figure}[h]
\centering
\subfloat[]{\includegraphics[width=0.55\linewidth]{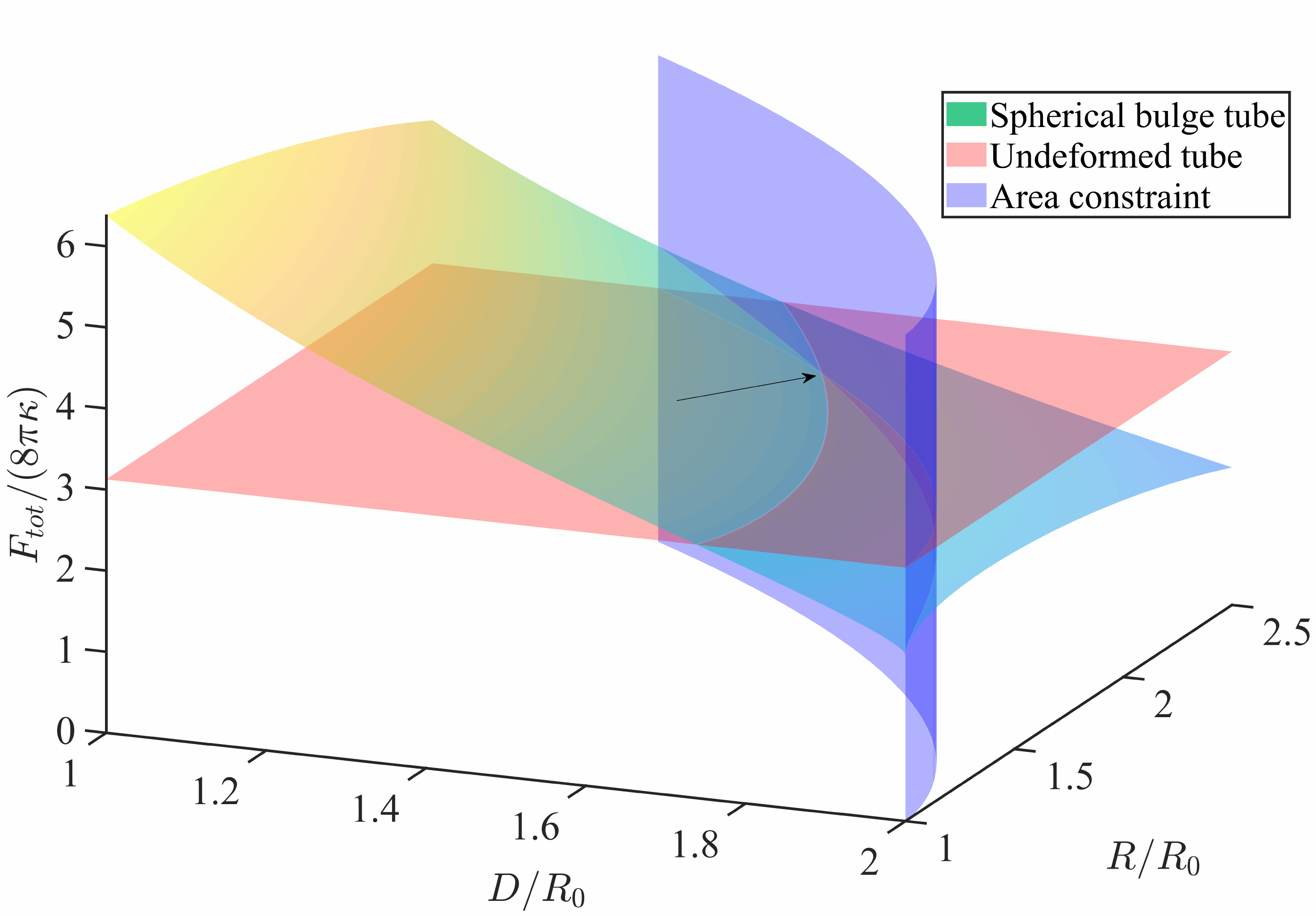}}
\subfloat[]{\includegraphics[width=0.45\linewidth]{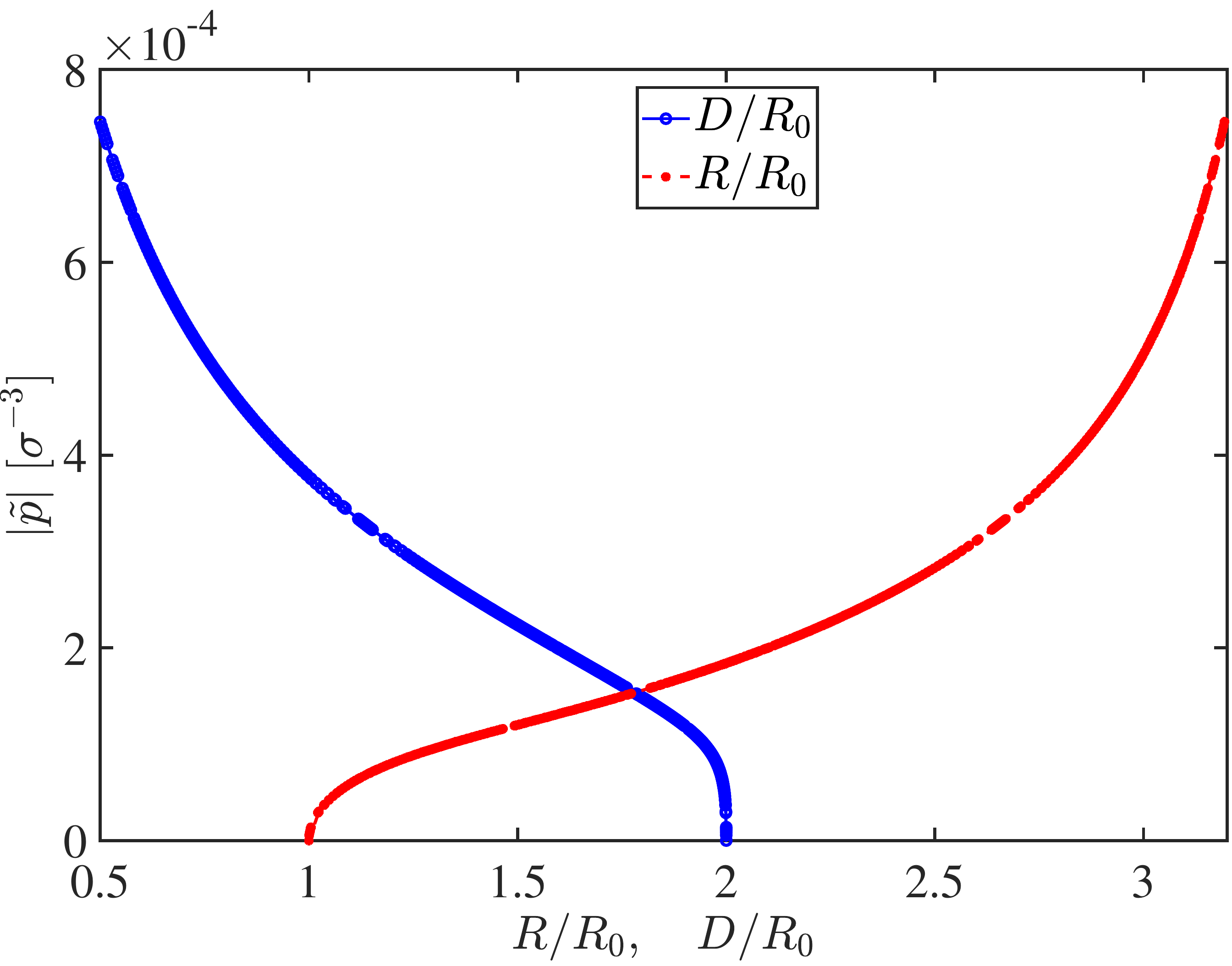}}
\caption{(a) Surface plot of total energy of the deformed cylinder acounting for the work done by the increase in internal pressure due to the insertion of the polymer, for $\Tilde{p} = 0.0002$ compared against the energy of an undeformed cylindrical tubule. Area constraint will give perticular value of $D/R_0$ and $R/R_0$ for that particular $\Tilde{p}$.  
(b) To get minimum energy of the deformed tubule, each $\Tilde{p}$ corresponds to particular solution of $D/R_0$ and $R/R_0$.}
\label{fig:pressure_D_R}
\end{figure}

Fig.~\ref{fig:pressure_D_R}(a) shows the intersection point of three planes formed by the elastic energies of the deformed and undeformed cylinder and the area as a function of the variational parameters $D/R_0$, and $R/R_0$. The intersection of all three planes yields the unique value of shape parameters for a given pressure $\tilde{p}$. Fig.~\ref{fig:pressure_D_R}(b), shows $\tilde{p}$ as a function of $R/R_0$, the radius of the spherical bulge and the diameter of the tube $D/R_0$. As shown in the figure, the equilibrium shape parameters $D/R_0 = 2$, and $D/R_0 = 1$ in the absence of pressure correspond to the undeformed cylinder (Fig.~\ref{fig:pressure_D_R}). The geometric parameters $R/R_0$ and $D/R_0$ values can be obtained for a particular choice $\Tilde{p}$, that minimises the elastic free energy obeying the constant area constraint of the deformed membrane tube.

\newpage

\subsection{\emph{Drop on a fibre} or sinusoidal bulge model}

Though the variational calculation using a ``cylindrical tubule with a spherical bulge'' ansatz gives a measure of the excess pressure exerted by the polymer it has a limitation \emph{i.e.} the kink in the deformed tubular membrane where the undeformed tubule joins with the deformed spherical bulge is unrealistic. We consider an ansatz where the deformed section smoothly connects to the undeformed tubule using a shape ansatz akin to a ``drop wetting a fibre''~\cite{b:degennes2013} given by, 
\begin{eqnarray}\label{eqn:2.1}
z = \frac{D}{2}+\frac{R}{2} \left[ 1 - \cos{\frac{2 \pi x}{\xi}}\right]~~~~~0\geq z \geq \xi,   
\end{eqnarray}
where $z(x)$ is the displacement of the membrane from its symmetry axis. The deformed axisymmetric shape smoothly connects with the undeformed segment, \emph{i.e.} a cylinder of length $L-\xi$ having a diameter $D$. In order to ensure the continuity of the membrane, we match the value of the function $z(x)$ and the derivative $z^{\prime}(x)$ at the intersection point between the deformed and undeformed segments. Further, we ensure that $z^{\prime}(x=\xi) = 0$. 

\begin{figure}[!h]
  \centering
  \subfloat[]{\includegraphics[width=0.43\linewidth]{SI06a.pdf}}
  \subfloat[]{\includegraphics[width=0.55\linewidth]{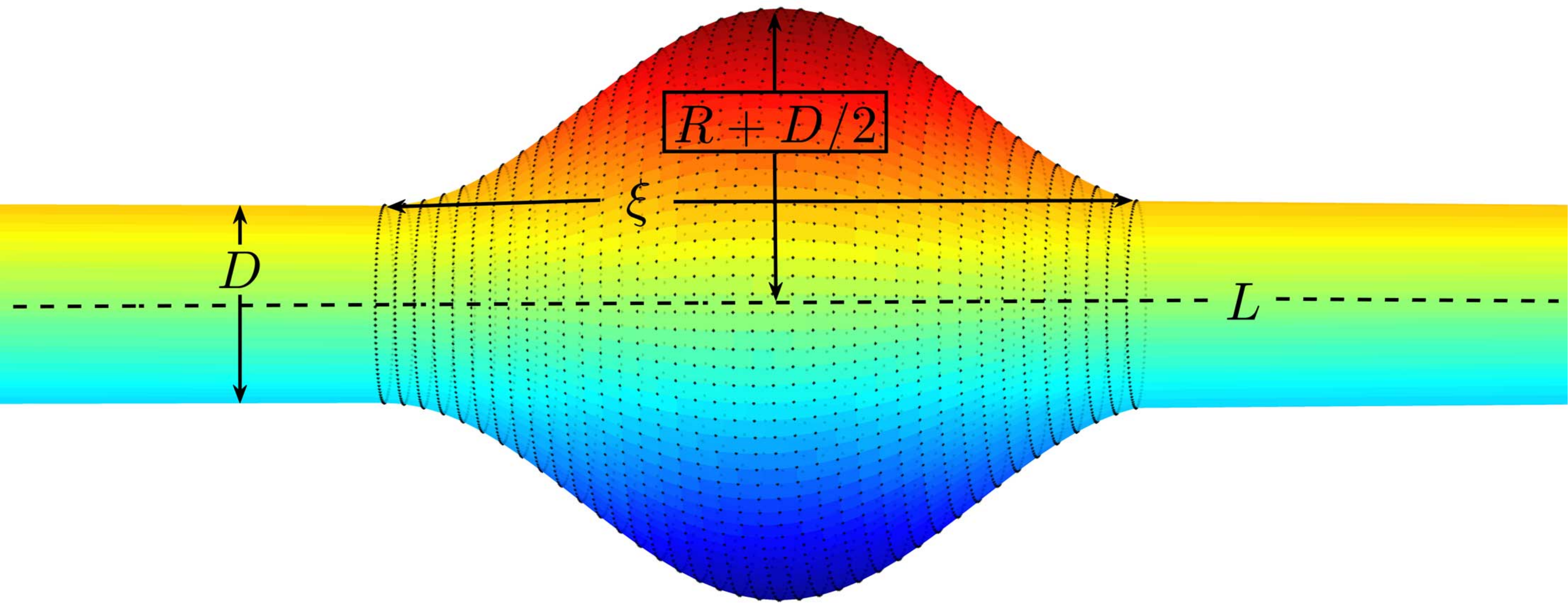}}
  \label{fig:sub2}
\caption{Schematic figure of a tubular membrane showing (a) an undeformed tube, and (b) a cylindrical section of diameter $D$ attached to a sinusoidal axisymmetric curve. The axial deformation scale of the bulge is $\xi$ while the maximum deformation of the membrane is $R + \frac{D}{2}$. Length of the unreformed and deformed cylinder is $L$.}
\label{fig:bulge_geometry_2}
\end{figure}

We compute the area of the deformed tube using the functional form of Eq.~(\ref{eqn:2.1}), where $dA = \sqrt{1+ \left(\frac{d z}{dx}\right)^2} dx$
\begin{eqnarray}\label{eqn:2.2}\nonumber
   A_{drop} &=&  \pi D(L-\xi) + \int_{0} ^{\xi} 2 \pi~ z~dA,\\
           &=&   \pi D(L-\xi) + \int_{0} ^{\xi} 2 \pi~ z~dA . 
\end{eqnarray}

Similarly the volume of the deformed tube using the functional form in Eq.~(\ref{eqn:2.1}) is given by,
\begin{eqnarray}\label{eqn:2.2}\nonumber
   V_{drop} &=&  \pi \frac{D^2}{4}(L-\xi) + \int_{0} ^{\xi} \pi~ z^2~dx,\\
           &=&   \pi \frac{D^2}{4}(L-\xi) +\frac{ \pi \xi}{8}(2 D^2 + 4 D R + 3 R^2) .
\end{eqnarray}

Using Fig.~(\ref{fig:bulge_geometry_2}(b)), principal curvatures of the meridian curve and the one perpendicular to the symmetry axis is given by, $c_1 = -\frac{\Ddot{z}}{(1+\Dot{z}^2)^{3/2}}$, and $c_2 = \frac{1}{z \sqrt{1+\Dot{z}^2}}$ (where $\Dot{z} = \frac{d z }{dx}$ and $\Ddot{z} = \frac{d^2 z }{dx^2}$) respectively. 

The elastic energy due to the deformation of the membrane can be calculated from the Helfrich-Canham Hamiltonian~\cite{p:deuling1976} and is given by, 
\begin{eqnarray}\label{eqn:1.2}\nonumber
  F_b  &=& \frac{\kappa}{2} \int_{A} (c_1 +c_2)^2 dA,\\\nonumber
       &=& 8 \pi \kappa \frac{L-\xi}{4D} + \kappa \pi \int_{0}^{\xi} \left(- \frac{\Ddot{z}}{(1+\Dot{z}^2)^{3/2}} +\frac{1}{z \sqrt{1+\Dot{z}^2}} \right)^2  z \sqrt{1+ \Dot{z}^2} ~dx ,\\
\frac{F_b}{8 \pi \kappa}&=&  \frac{L-\xi}{4D} + \frac{1}{8} \int_{0}^{\xi} \left(- \frac{\Ddot{z}}{(1+\Dot{z}^2)^{3/2}} +\frac{1}{z \sqrt{1+\Dot{z}^2}} \right)^2  z \sqrt{1+ \Dot{z}^2} ~dx. 
\end{eqnarray}

The work done by the pressure in deforming the cylindrical tubule to bring about a shape change is given by Eq.~\ref{eqn:1.5}

\begin{eqnarray}\label{eqn:1.6}\nonumber
F_p &=& -\Delta p \left[ \pi \frac{D^2}{4}(L-\xi) +\frac{ \pi \xi}{8}(2 D^2 + 4 D R + 3 R^2)\right],\\\nonumber
\frac{F_p}{8 \pi \kappa} &=& -\Tilde{p} \left[ \frac{D^2}{32}(L-\xi) +\frac{\xi}{64}(2 D^2 + 4 D R + 3 R^2)\right],\\
\end{eqnarray}
where $\Tilde{p} = \Delta p/\kappa$. 

Fig.~\ref{fig:pressure_D_R} shows the isosurface of the elastic energy of the deformed tubule using the drop on a fibre shape ansatz, and the undeformed cylindrical tubule plotted in the $D/R_0$ - $\tilde{R}/R_0$ parametric plane ($\tilde{R} = R + D/2$), for a fixed value of $\xi$. As in the case for the spherical bulge model the locus of the intersection points between the two isosurfaces traces out parameter values for which the deformed cylinder has the same energy as an undeformed tubule. Fig.~\ref{fig:pressure_D_R}(b) shows a family of curves for different values of $\xi$ and a given value of shape parameters $\tilde{R}/R_0$, and $D/R_0$, that minimise the elastic free energy. 

\begin{figure}[h]
\centering
  \subfloat[]{\includegraphics[width=0.38\linewidth]{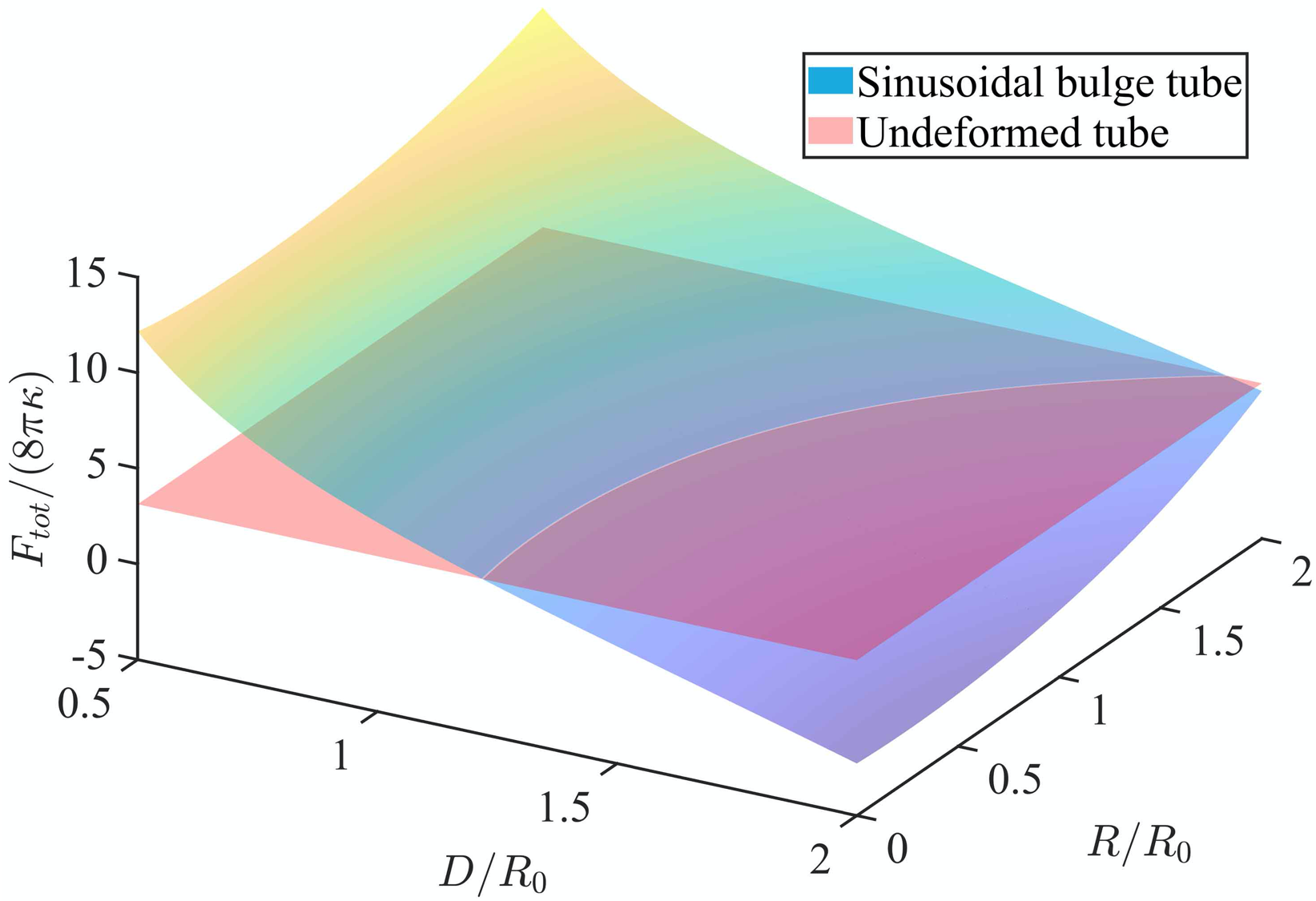}\label{13a}}
  \subfloat[]{\includegraphics[width=0.3\linewidth]{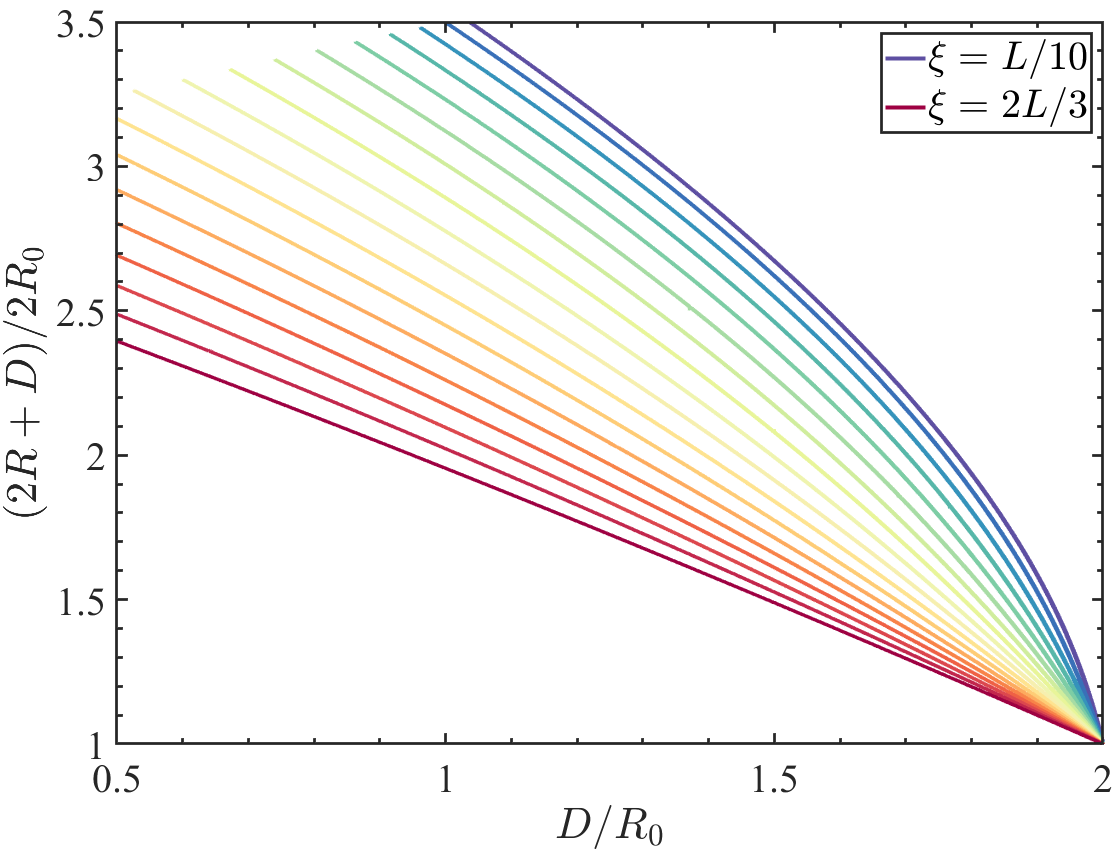}\label{13b}}
  \subfloat[]{\includegraphics[width=0.32\linewidth]{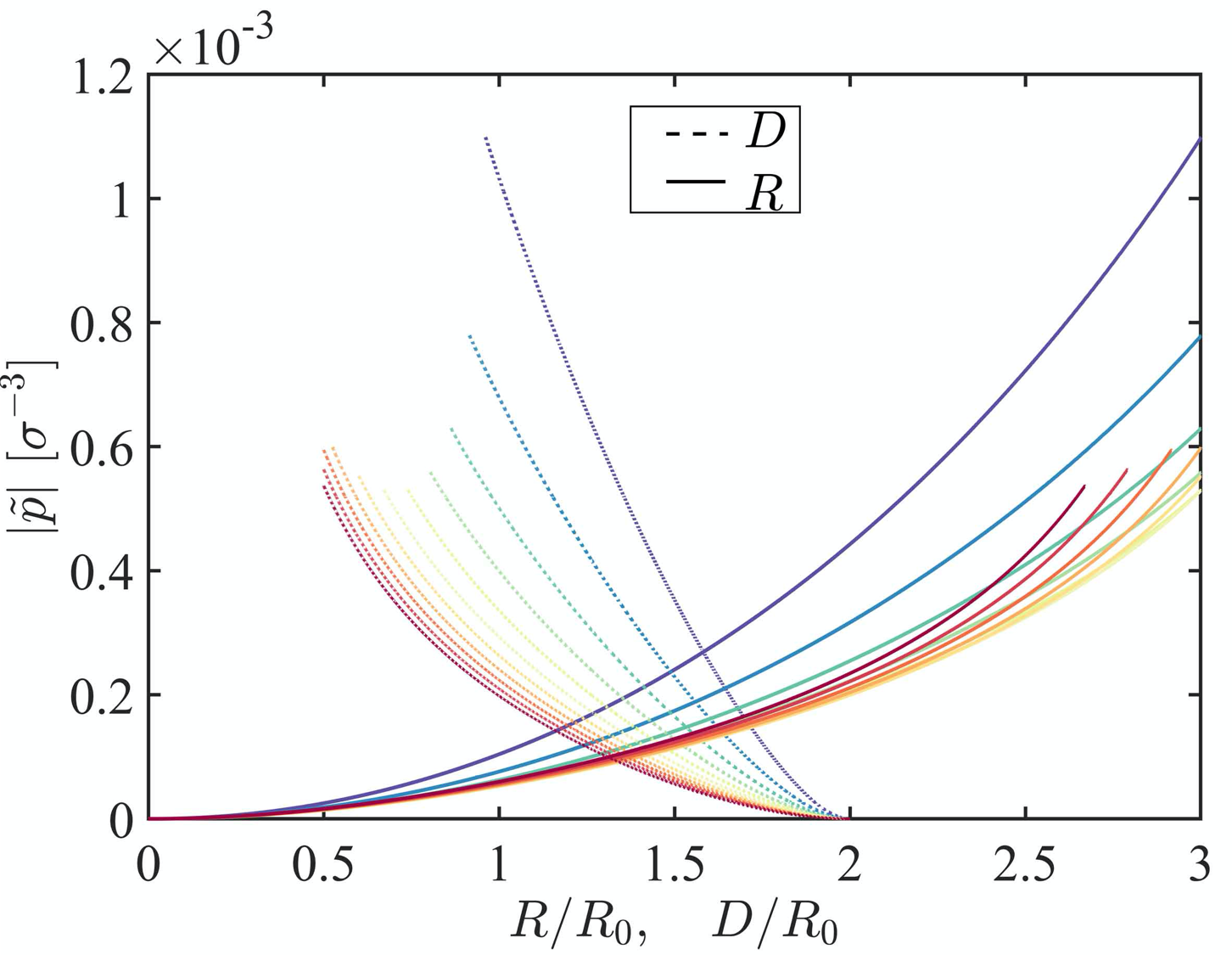}\label{13c}}
\caption{(a) Total energy of the deformed ``drop on a fibre'' geometry for $\Tilde{p} = 0.0002$, $R_0 = 12 $, $L =300$ and undeformed cylindrical tube. Two surfaces intersect at a point where the total energy of the deformed tubule equals that of the elastic energy of the undeformed tube. (c) $\Tilde{p}$ corresponding $D/R_0$ and $R/R_0$ with particular $\xi$ (value used in the (b)) to get minimum energy of the deformed tube.}
\label{fig:si_13}
\end{figure}
In conjunction with the area conservation constraint imposed on the deformed surface we obtain a unique value of shape parameters $D/R_0$ and $R/R_0$ for which the elastic energies of the deformed and undeformed surfaces are equal for a given value of $\xi$. The family of curves for which the area constraint is satisfied is shown in Fig.~\eqref{13b}. The intersection of the family of curves that satisfy equal energy and area constraint between deformed and undeformed tubules gives rise to different pressures $\tilde{p}$. 

We note that the three parameter variational minimisation of the elastic free energy is non-unique, as the number of variables $\xi$, $\tilde{R}/R_0$, and $D/R_0$ is more than the number of constraints. It is thus possible to find more than one solution with a different combination of $D/R_0$, $\tilde{R}/R_0$, and $\xi$ that simultaneously minimise the free energy and satisfy the constant area constraint. We therefore need an additional input to uniquely determine the shape and hence the pressure exerted by an encapsulated polymer chain. This can be carried out in imaging studies by measuring the axial deformation length $\xi$. One can than use the ``drop on a fibre'' shape ansatz and the calculational scheme outlined here to uniquely determine the pressure exerted by a semiflexible chain on the walls of a soft tubule. Along with the calculation of pressure based on free energy of the confined polymer this forms a self-consistent scheme to connect the conformational states of confined chains with their thermodynamic behavior.

\section{SI: Movie}
SI movie shows the time evolution of a polymer encapsulated in a tubule starting from a random initial polymer conformation to the equilibrium sate obtained using CGMD simulations. The first and second columns show the axial and radial cross sections of lipid tubule. The encapsulated semiflexible polymer (green beads) of size $N=2000 \sigma$, is confined in the bilayer lipid tubule having hydrophilic head groups (red beads) and hydrophobic tails (blue beads). For rigid tubules having $\kappa = 24 k_{B} T $ (a) swollen chain is seen (first row), while for softer $\kappa = 12 k_B T$ tubules (second row) (b) a prolate ellipsoidal conformation is observed. The persistence length of the polymer in this case is $l_p = 13 \sigma$. The third row (c) shows a toroidal coil for a polymer with $l_{p} \simeq 200 \sigma$, (such that $l_{p} \gtrsim \xi$).
 
\vspace{0.5cm}

\end{document}